\documentclass[runningheads]{svmult}

\usepackage{makeidx}   
\usepackage{graphicx}  
\usepackage{subeqnar}  
\usepackage{multicol}  
\usepackage{physprbb}  
\makeindex             

\newcommand{\EQ}{\begin{equation}}
\newcommand{\EN}{\end{equation}}
\newcommand{\EQA}{\begin{eqnarray}}
\newcommand{\ENA}{\end{eqnarray}}
\newcommand{\eq}[1]{(\ref{#1})}
\newcommand{\EEq}[1]{Equation~(\ref{#1})}
\newcommand{\Eq}[1]{Eq.~(\ref{#1})}

\newcommand{\Sec}[1]{Sect.~\ref{#1}}
\newcommand{\Secs}[2]{Sects~\ref{#1} and \ref{#2}}
\newcommand{\Fig}[1]{Fig.~\ref{#1}}
\newcommand{\FFig}[1]{Figure~\ref{#1}}

\newcommand{\Figs}[2]{Figs~\ref{#1} and \ref{#2}}

\newcommand{\bra}[1]{\langle #1\rangle}

\newcommand{\mean}[1]{\overline #1}
\newcommand{\meanemf}{\overline{\mbox{\boldmath ${\cal E}$}} {}}
\newcommand{\meanAA}{\overline{\bf{A}}}
\newcommand{\meanB}{\overline{B}}
\newcommand{\meanBB}{\overline{\bf{B}}}
\newcommand{\meanJJ}{\overline{\bf{J}}}

{}
{}
{}
{}
{}
{}
{}
\newcommand{\xx}{\mbox{\boldmath $x$} {}}

\newcommand{\uu}{{\bf{u}}}
\newcommand{\BB}{{\bf{B}}}
\newcommand{\JJ}{{\bf{J}}}
\newcommand{\jj}{{\bf{j}}}
\newcommand{\AAA}{{\bf{A}}}
\newcommand{\aaaa}{{\bf{a}}}
\newcommand{\bb}{{\bf{b}}}
\newcommand{\nn}{\mbox{\boldmath $n$} {}}

\newcommand{\ff}{\mbox{\boldmath $f$} {}}

\newcommand{\EE}{{\bf{E}}}

\newcommand{\kk}{\mbox{\boldmath $k$} {}}
\newcommand{\SSS}{{\bf{S}}}

\newcommand{\nab}{\vec{\nabla}}

\newcommand{\oo}{\vec{\omega}}

\newcommand{\ii}{{\rm i}}

\newcommand{\DD}{{\rm D} {}}
\newcommand{\dd}{{\rm d} {}}

\def\la{\mathrel{\mathchoice {\vcenter{\offinterlineskip\halign{\hfil
$\displaystyle##$\hfil\cr<\cr\sim\cr}}}
{\vcenter{\offinterlineskip\halign{\hfil$\textstyle##$\hfil\cr<\cr\sim\cr}}}
{\vcenter{\offinterlineskip\halign{\hfil$\scriptstyle##$\hfil\cr<\cr\sim\cr}}}
{\vcenter{\offinterlineskip\halign{\hfil$\scriptscriptstyle##$\hfil\cr<\cr\sim\cr}}}}}
\def\ga{\mathrel{\mathchoice {\vcenter{\offinterlineskip\halign{\hfil
$\displaystyle##$\hfil\cr>\cr\sim\cr}}}
{\vcenter{\offinterlineskip\halign{\hfil$\textstyle##$\hfil\cr>\cr\sim\cr}}}
{\vcenter{\offinterlineskip\halign{\hfil$\scriptstyle##$\hfil\cr>\cr\sim\cr}}}
{\vcenter{\offinterlineskip\halign{\hfil$\scriptscriptstyle##$\hfil\cr>\cr\sim\cr}}}}}
\newcommand{\ea}{{\rm et al.\ }}

\def\half{{\textstyle{1\over2}}}

\def\onethird{{\textstyle{1\over3}}}

\newcommand{\s}{\,{\rm s}}

\newcommand{\cm}{\,{\rm cm}}

\newcommand{\km}{\,{\rm km}}

\newcommand{\Mm}{\,{\rm Mm}}
\newcommand{\Mx}{\,{\rm Mx}}

\newcommand{\yr}{\,{\rm yr}}

\newcommand{\yjgr}[3]{: {J. Geophys. Res.} {#2}, #3 (#1)}
\newcommand{\yapj}[3]{: {Astrophys. J.} {#2}, #3 (#1)}
\newcommand{\yapjl}[3]{: {Astrophys. J.} {#2}, #3 (#1)}

\newcommand{\yan}[3]{: {Astron. Nachr.} {#2}, #3 (#1)}

\newcommand{\yana}[3]{: {Astron. Astrophys.} {#2}, #3 (#1)}

\newcommand{\ygafd}[3]{: {Geophys. Astrophys. Fluid Dyn.} {#2}, #3 (#1)}
\newcommand{\yjfm}[3]{: {J. Fluid Mech.} {#2}, #3 (#1)}
\newcommand{\ypf}[3]{: {Phys. Fluids} {#2}, #3 (#1)}
\newcommand{\ypp}[3]{: {Phys. Plasmas} {#2}, #3 (#1)}

\newcommand{\yphy}[3]{: {Physica} {#2}, #3 (#1)}

\newcommand{\yprl}[3]{: {Phys. Rev. Lett.} {#2}, #3 (#1)}
\newcommand{\yphl}[3]{: {Phys. Lett.} {#2}, #3 (#1)}

\newcommand{\ymn}[3]{: {Month. Not. Roy. Astron. Soc.} {#2}, #3 (#1)}
\newcommand{\ynat}[3]{: {Nature} {#2}, #3 (#1)}

\newcommand{\ysph}[3]{: {Solar Phys.} {#2}, #3 (#1)}
\newcommand{\ypr}[3]{: {Phys. Rev.} {#2}, #3 (#1)}

\newcommand{\yjour}[4]{: {#2} {#3}, #4 (#1)}

\newcommand{\ybook}[3]{: {\it #2} (#3 #1)}
\newcommand{\yproc}[7]{: `#4'. In: {\em #5}, ed. by #6 (#7 #1) pp.\ #2-#3}

\newcommand{\pgafd}[1]{: {Geophys. Astrophys. Fluid Dyn.}, in press (#1)}

\newcommand{\papj}[1]{: {Astrophys. J.}, in press (#1)}

\usepackage{html}
\usepackage{url}

\begin{document}
\title*{The helicity issue in large scale dynamos}
\toctitle{The helicity issue in large scale dynamos}
\titlerunning{The helicity issue in large scale dynamos}
%
\author{Axel Brandenburg}
\authorrunning{Axel Brandenburg}
\institute{NORDITA, Blegdamsvej 17, DK-2100 Copenhagen \O, Denmark\\
\today,~ $ $Revision: 1.29 $ $}

\maketitle              

\begin{abstract}
The connection between helically isotropic MHD turbulence and mean-field
dynamo theory is reviewed. The nonlinearity in the mean-field theory is not yet well
established, but detailed comparison with simulations begin to help select
viable forms of the nonlinearity. The crucial discriminant is
the magnetic helicity, which is known to evolve only on a slow resistive
time scale in the limit of large magnetic Reynolds number.
Particular emphasis is put
on the possibility of memory effects, which means that an additional explicitly
time-dependent equation for the nonlinearity is solved simultaneously
with the mean-field equations. This approach leads to better agreement
with the simulations, while it would also produce more favorable
agreement between models and stellar dynamos.
\end{abstract}

\section{Introduction}

In an early paper Parker \cite{Par55} identified cyclonic
convection as a key process for converting large scale toroidal magnetic
field into poloidal fields that have coherence over about half a
hemisphere. This process is now generally
referred to as the $\alpha$-effect, although it may arise not only
from thermal buoyancy \cite{SKR66}, but also
from magnetic buoyancy \cite{Lei69,FMSS94,BS98},
the magneto-rotational instability \cite{BNST95,BD97},
or some other magnetic instability \cite{GilFox97}.
In each case the effect of the Coriolis force together
with some kind of radial stratification is crucial for making the
motions helical \cite{KR80}.
Upward moving fluid expands, and the Coriolis force makes it rotate
retrograde, causing negative (positive) kinetic helicity in the northern
(southern) hemisphere. Downward moving fluid contracts, rotates in
the prograde direction and contributes in the same sense to negative
(positive) kinetic helicity on the northern (southern) hemisphere.
This causes a positive $\alpha$-effect in the northern hemisphere,
but if magnetic stresses and shear become strong (for example in
accretion discs) the sign may reverse \cite{Bra98,RP00}.

When combined with differential rotation, the main outcome
of $\alpha$-effect models is the possibility of cyclic magnetic fields
associated with latitudinal migration. The first global (two-dimensional)
models were presented by Steenbeck \& Krause \cite{SK69}, but similar models,
with different physics, are still being studied today \cite{RB95,DC99,KRS01}.
The migratory behavior is best seen in contours
of the longitudinally averaged mean magnetic field versus latitude and
time, which should show tilted structures converging to the equator. Such
plots can be compared with the solar butterfly diagram of sunspot numbers
(so called because the structures resemble a sequence of butterflies).

A key property of all these models is that not only the motions are
helical, but the large scale magnetic field itself is also helical.
Of course, not all dynamos require helicity, but nonhelical dynamos tend
to generate preferentially small-scale fields \cite{Cat99}. In a recent attempt,
Vishniac \& Cho \cite{VC01} proposed a mechanism relevant in particular to accretion discs
where shear is strong. Their mechanism would not lead to the production
of net magnetic helicity, but numerical simulations \cite{AB01}
failed so far to show convincingly {\it large scale} dynamo action based
on the proposed mechanism. Shear does produce
large scale fields, but only in the toroidal direction. It does
not explain the latitudinal coherence of the field over several
tens of degrees (corresponding to several hundred megameters).
On the other hand, there is direct observational evidence that the solar
magnetic field is indeed helical. (We shall return to observations
in \Sec{Ssun}.)

The trouble with helical fields is that magnetic helicity is conserved
by the induction equation in the ideal limit
and can only change on a resistive time scale,
provided there is no significant loss through boundaries (at the surface
or the equator, for example). This
approximate magnetic helicity conservation leads to
magnetic field saturation on a resistive time scale \cite{B01}. Depending
on how effectively the boundaries transmit magnetic energy and helicity,
the final saturation amplitude will be lowered if losses occur
preferentially on large scales while the (linear) growth
rate of magnetic energy (past initial saturation) remains {\it unchanged}
\cite{BD01}. In this sense final saturation can be achieved earlier.
The above results are particularly clear when the flow is nearly fully
helical, i.e.\ when the normalized kinetic helicity,
$\epsilon_{\rm f}\equiv\bra{\oo\cdot\uu}/(\omega_{\rm rms}u_{\rm rms})$,
where $\oo=\nab\times\uu$ is the vorticity, is large.
In the sun, and probably in all other celestial bodies with rotating
turbulence, the relative kinetic helicity is small;
$\epsilon_{\rm f}\sim5\%$.
It is however
this small helical fraction of the turbulence
that is responsible for the a finite but small $\alpha$-effect, so
a proper understanding of its dynamics is crucial if one wants to build
models based on the $\alpha$-effect.
Below we shall also discuss the alternative that astrophysical dynamos
may shed preferentially small scale helical fields through the boundaries.
This could theoretically enhance large scale dynamo action
\cite{BF00,KMRS00,BDS02}.

\section{Magnetic helicity production}
\label{Ssun}

Before we begin discussing the magnetic helicity problem and possible
remedies we need to be sure that the solar dynamo is indeed likely to involve
significant amounts of magnetic helicity. There is direct observational evidence
that the field of the sun is actually helical. Firstly, active regions
are known to have systematically different signs of current helicity
in the two hemispheres \cite{See90,PCM95,BZAZ99,PL00}; preferentially
negative (positive)
in the northern (southern) hemisphere. Secondly, magnetic helicity
flux from the solar surface has also been inferred and this confirms the same
sign as that of the current helicity.
The magnetic helicity flux driven by the surface differential
rotation has been estimated by Berger \& Ruzmaikin \cite{BR00}, who find
a total magnetic helicity flux on the order of $4\times10^{46}\Mx^2$ over
the 22 year solar cycle. Similar values were also found by
DeVore \cite{DeV00}. Finally, Chae \cite{Chae00} estimated
the magnetic helicity flux based on counting the crossings of pairs of
flux tubes.  Combined with the assumption that two nearly aligned flux
tubes are nearly parallel (rather than antiparallel) his results again
suggest that the magnetic helicity is preferentially negative (positive)
in the northern (southern) hemisphere.

The magnetic helicity is noisy, i.e.\ the sign can fluctuate and has
only on average systematic behavior. This reflects the fact that only
a fraction of the turbulence is helical. Thus, detailed understanding
and measurements of the departures from systematic behavior is just
as important as understanding and recording the systematic behavior.

In astrophysical flows, kinetic helicity can be generated in rotating
stratified turbulence \cite{KR80}. Such flows are
intrinsically anisotropic. Whilst this is not a problem for numerical
simulations, it definitely complicates the theoretical understanding
and one should not be surprised if some fundamental aspects of
mean-field theory (e.g.\ $\alpha$ proportional to $-\bra{\oo\cdot\uu}$)
are not recovered. We just mention that under certain conditions,
stratified rotating flows exhibit an $\alpha$-tensor whose vertical
($z$-$z$) component has the opposite sign as the horizontal
($x$-$x$ and $y$-$y$) components \cite{BNPST90,Fer92,RK93,OSB01}.)

In the following we concentrate on the isotropic helical aspects of the
turbulence. This is accomplished by driving the flow with random
polarized waves in a periodic domain.
In most of the cases we use fully helical turbulence,
but in many estimates the fraction of helicity enters only as an
additional scaling factor \cite{BDS02}.
The main goal here is a better understanding of
the $\alpha$-effect and the nonlinear feedback when the field becomes
dynamically important. We therefore discuss in detail the perhaps simplest
possible system: the $\alpha^2$-dynamo in a periodic box.

In spherical geometry,
the term $\alpha^2$-dynamo refers to the fact that both large scale poloidal
and toroidal fields are maintained against ohmic decay by the
$\alpha$-effect. By contrast, the $\alpha\Omega$-dynamo is one where
the large scale toroidal field is generated mostly by differential rotation
(the $\Omega$-effect) and the $\alpha$-effect can be neglected by
comparison. If the $\alpha$-effect is not neglected one speaks of
an $\alpha^2\Omega$-dynamo.
We stress that the $\alpha^2$-dynamo has nothing to do with the so-called
small scale dynamo. These are turbulent dynamos operating only on scales
less than the energy-carrying scale of the turbulence.  They are quite
common if the flows are non-helical. By contrast, both $\alpha^2$ and
$\alpha\Omega$-dynamos also generate fields on large scales, but they
are necessarily accompanied by some level of small scale fields as well.

In its simplest form the $\alpha^2$-dynamo equations for isotropic
$\alpha$ and turbulent diffusivity $\eta_{\rm t}$ can be written as
\EQ
{\partial\mean\BB\over\partial t}=
\nab\times\left(\alpha\meanBB-\eta_{\rm T}\mu_0\meanJJ\right),
\label{simplest_dyn_eqn}
\EN
where $\meanJJ=\nab\times\meanBB/\mu_0$ is the mean current density and $\mu_0$
is the magnetic permeability. This equation permits plane wave solutions
of the form $\meanBB=\hat{\BB}\exp(\lambda t+\ii\kk\cdot\xx)$, with the
dispersion relation $\lambda=\pm|\alpha|k-\eta_{\rm T}k^2$, where $k=|\kk|$.
The maximum of $\lambda$ is where $\dd\lambda/\dd k=0$, which yields
\EQ
k=k_{\rm max}=\alpha/(2\eta_{\rm T}).
\label{kmax}
\EN
For a periodic box of size $L^3$,
the most easily excited mode has $k=k_1\equiv2\pi/L$, and the $\kk$ vector can
point in any of the three coordinate directions. For $\alpha<0$
(the case considered in Ref.~\cite{B01}), the three
possible eigenfunctions are
\EQ
\BB^{(z)}=B_0\!\pmatrix{\cos k_1 z\cr\sin k_1 z\cr0}\!,\quad
\BB^{(x)}=B_0\!\pmatrix{0\cr\cos k_1 x\cr\sin k_1 x}\!,\quad
\BB^{(y)}=B_0\!\pmatrix{\sin k_1 y\cr0\cr\cos k_1 y}\!,
\EN
where we have ignored arbitrary phase shifts in any of the three directions.
All three solutions have been found in the simulations \cite{B01}.

In the simulations there is of course no explicit $\alpha$-effect in the
usual sense, because we just solve the primitive MHD equations.
The turbulence does, however, display {\it collective behavior} -- just as
it is expected based on mean-field $\alpha^2$-dynamo theory, as explained
above. We begin with the simulations.

\section{Helical turbulence: prototype of an $\alpha^2$-dynamo}
\label{Shelical}

We consider a compressible isothermal gas with constant sound speed
$c_{\rm s}$, constant dynamical viscosity $\mu$, and constant magnetic
diffusivity $\eta$. To make
sure the magnetic field stays solenoidal, i.e.\ $\nab\cdot\BB=0$, we
express $\BB$ in terms of the magnetic vector potential $\AAA$, so the
field is written as $\BB=\nab\times\AAA$. The governing equations for
density $\rho$, velocity $\uu$, and magnetic vector potential $\AAA$,
are given by
\EQ
{\DD\ln\rho\over\DD t}=-\nab\cdot\uu,
\label{dlnrhodt}
\EN
\EQ
{\DD\uu\over\DD t}=-c_{\rm s}^2\nab\ln\rho+{\JJ\times\BB\over\rho}
+{\mu\over\rho}(\nabla^2\uu+\onethird\nab\nab\cdot\uu)+\ff,
\label{dudt}
\EN
\EQ
{\partial\AAA\over\partial t}=\uu\times\BB-\eta\mu_0\JJ-\nab\phi,
\label{dAdt}
\EN
where ${\rm D}/{\rm D}t=\partial/\partial t+\uu\cdot\nab$ is the advective
derivative. The current density, $\JJ=\nab\times\BB/\mu_0$, is obtained
in the form $\mu_0\JJ=-\nabla^2\AAA+\nab\nab\cdot\AAA$.  We often use
$\phi=0$ as a convenient gauge for the electrostatic potential. Other
frequent choices are $\phi=-\eta\nab\cdot\AAA$, $\phi=\uu\cdot\AAA$, or
combinations of these \cite{B01b}.
The Coulomb gauge, $\nab\cdot\AAA=0$, corresponds
to $\phi=-\nab\cdot\EE$, where $\EE=-\uu\times\BB+\eta\mu_0\JJ$, but the
original reason for solving for $\AAA$ instead of $\BB$ was just to get
rid of the solenoidality condition, so the Coulomb condition has
computational disadvantages.

For the following it is useful to recall that each vector field can be
decomposed into a solenoidal and two vortical parts with positive and
negative helicity, respectively. These are also referred to as
Chandrasekhar-Kendall functions. Although it is often useful to decompose
the magnetic field into positive and negative helical parts, we also use
the helical fields (with positive helicity) as forcing function $\ff$
of the flow. We restrict ourselves to functions selected from a finite
band of wavenumbers around the wavenumber $k_{\rm f}$, but direction
and amplitude
are chosen randomly at each timestep. Details can be found in Ref.~\cite{B01}.
Similar work was first carried out
by Meneguzzi \ea \cite{MFP81}, but at the time one was barely able
to run even until
saturation. Throughout the nineties, work has been done on forced ABC flows
\cite{GS91,GSG91,BP99}.
In none of these investigations, however, the issue of resistively slow magnetic
helicity evolution past initial saturation has been noted. It is
exactly this aspect that has now become so crucial in understanding the
saturation behavior of nonlinear dynamos. We begin by discussing first
the linear (kinematic) evolution of the magnetic field.

\subsection{Linear behavior}

If the magnetic Reynolds number, defined here as
$R_{\rm m}=u_{\rm rms}/(\eta k_{\rm f})$, exceeds a certain critical
value, $R_{\rm m}^{\rm(crit)}$, there is dynamo action. For helical flows, the
so defined $R_{\rm m}^{\rm(crit)}$ is between 1 and 2 (see Table~1 of
Ref.~\cite{B01}). In the supercritical
case, $R_{\rm m}>R_{\rm m}^{\rm(crit)}$, the field grows exponentially
with growth rate $\lambda$, which scales with the inverse turnover time,
$u_{\rm rms}/k_{\rm f}$. The resistively limited saturation behavior that
will be discussed below in full detail has no obvious correspondence
in the kinematic stage when the field is too weak to affect the motions
by the Lorentz force \cite{{Gil02}}.
Nevertheless, there is actually a subtle effect
on the shape of the eigenfunction as $R_{\rm m}$ increases. Before
we can appreciate this, we need to discuss the effect the kinetic helicity
has on the field.

A helical velocity tends to drive helicity in the magnetic field as well,
but in the nonresistive limit magnetic helicity conservation dictates
that $\bra{\AAA\cdot\BB}=\mbox{const}=0$ if the initial field (or at
least its helicity) was infinitesimally weak.
Thus, there must be some kind of magnetic helicity
cancelation. Under homogeneous isotropic conditions there cannot be
a spatial segregation in positive and negative helical parts.
Instead, there is a spectral segregation:
there is a bump at the forcing wavenumber and another `secondary'
bump at somewhat smaller wavenumber. The two bumps have opposite
sign of magnetic helicity such that the net magnetic helicity
is zero. At the forcing wavenumber, the sign of magnetic helicity
agrees with that of the kinetic helicity, but at smaller wavenumber
the sign of magnetic helicity is opposite. At small $R_{\rm m}$,
this secondary peak can be identified with the wavenumber where the
corresponding $\alpha^2$-dynamo has maximum growth rate; see \Eq{kmax}.
Simulations show that as $R_{\rm m}$ increases, $k_{\rm max}$ approaches
$\half k_{\rm f}$ \cite{BDS02}.
This agrees qualitatively
with earlier results \cite{Gil02,HCK96} which suggested
that the magnetic helicity approaches zero in
the high-$R_{\rm m}$ limit.

\begin{figure}[h!]\centering\includegraphics[width=0.95\textwidth]{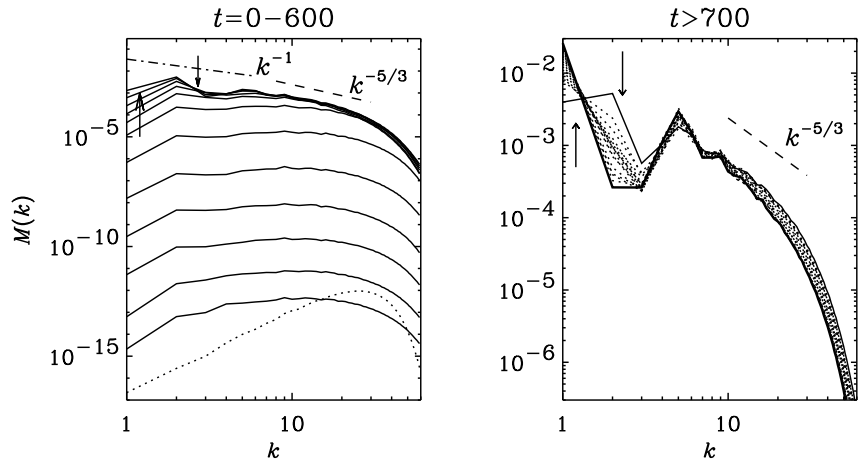}\caption{
Power spectra of magnetic energy of Run~3 of Ref.~\cite{B01}. During the initial
growth phase the field saturates at small scales first and only later at
large scales (left hand panel). Later, when also the large scale field
saturates, the field at intermediate scales ($k=2$, 3, and 4) becomes
suppressed. In the second panel, intermediate times are shown as dotted
lines, $t=700$ is shown in solid and $t=1600$ is shown as a thick solid
line. The forcing wavenumber is $k_{\rm f}=5$.
}\label{Fpspec_growth_passot}\end{figure}

\subsection{Nonlinear behavior}

\begin{figure}[h!]\centering\includegraphics[width=0.95\textwidth]{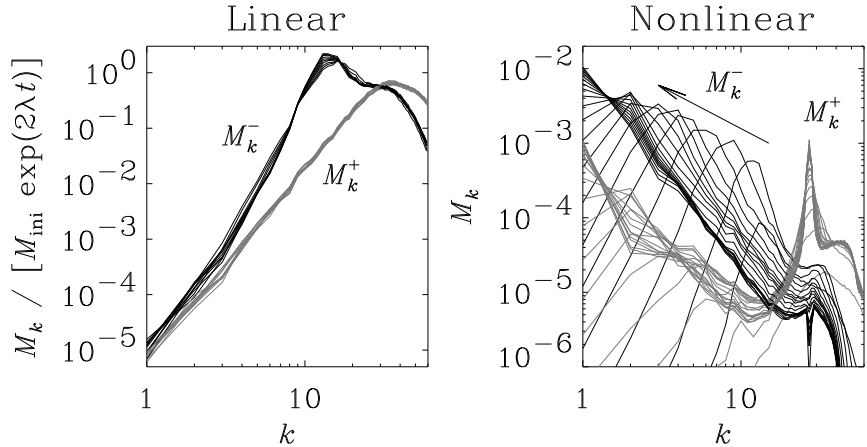}\caption{
Power spectra of magnetic energy of positively and negatively polarized
parts ($M_k^+$ and $M_k^-$)
in the linear and nonlinear regimes. The spectra in the linear
regime have been compensated by the exponential growth factor to make
them collapse on top of each other. Here the forcing wavenumber is in
the dissipative subrange, $k_{\rm f}=27$, but this allows enough scale
separation to see the inverse transfer of magnetic energy to smaller $k$.
The data are from Run~B of Ref.~\cite{BS02}.
}\label{Fpspec_pm_satkin}\end{figure}

\begin{figure}[h!]\centering\includegraphics[width=0.95\textwidth]{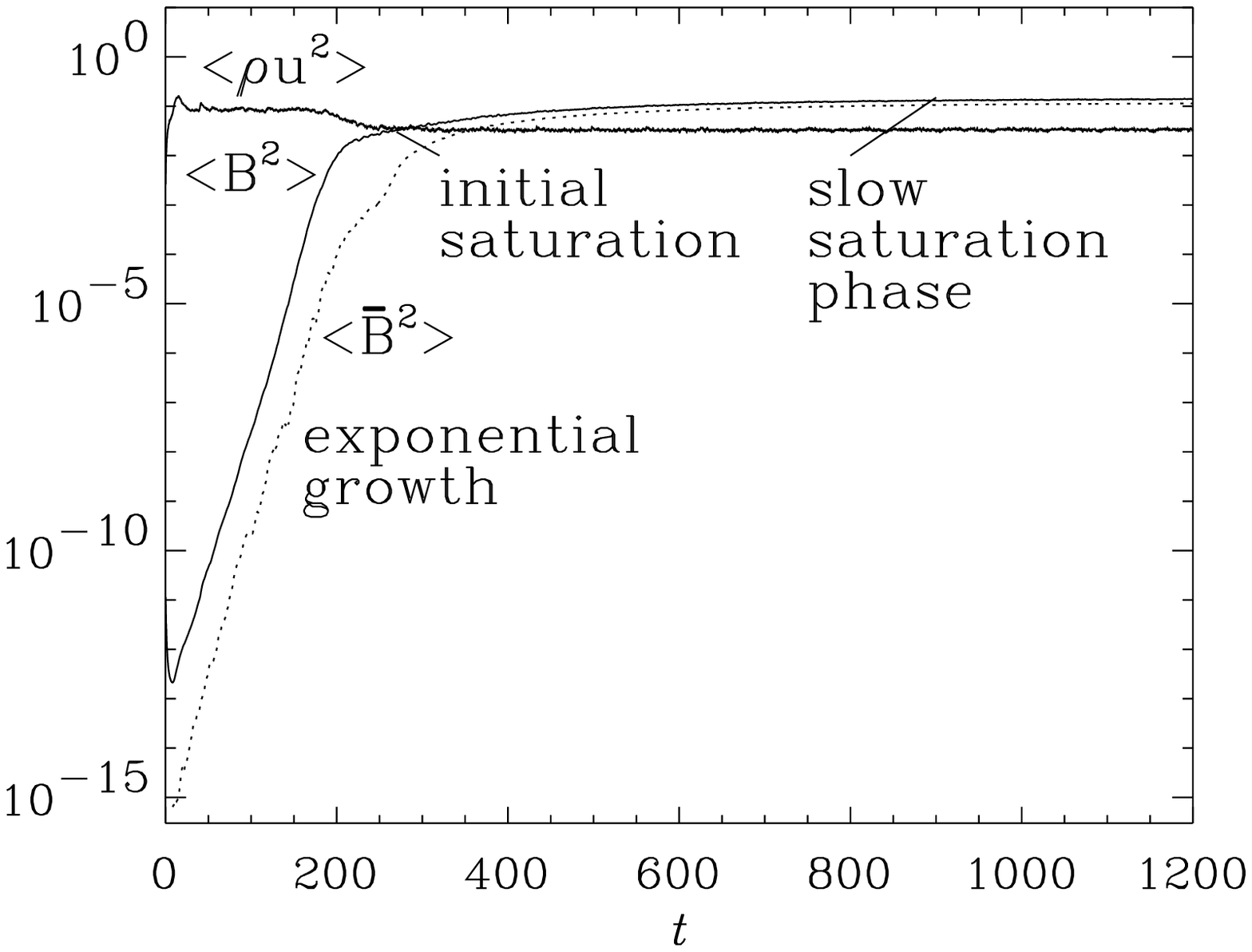}\caption{
The three stages of the magnetic field growth: exponential growth
until initial saturation (when $\bra{\BB^2}/\mu_0=\bra{\rho\uu^2}$),
followed by a (resistively) slow saturation phase. In this plot we
have used $\mu_0=1$. The energy of the large scale magnetic field,
$\bra{\meanBB^2}$, is shown for comparison. The data are from Run~3
of Ref.~\cite{B01}.
}\label{Fpjbm_decay_nfit_run3passot}\end{figure}

Eventually, the magnetic energy stops increasing exponentially. This is
due to the nonlinear terms, in particular the Lorentz force $\JJ\times\BB$
in \Eq{dudt}, which begins to affect the velocity field. The temporal
growth of the power spectra saturates, but only partially; see
\Fig{Fpspec_growth_passot}, where we show data from a run with
forcing at wavenumber $k_{\rm f}=5$.
In the left hand panel we see that by the time $t=600$ the power spectra
have saturated at larger wavenumbers, $k\ga3$. It takes until $t\simeq1600$
for the power spectra to be saturated also at $k=1$ (right hand panel
of \Fig{Fpspec_growth_passot}). In order to see more clearly the behavior
at large scales, we show in \Fig{Fpspec_pm_satkin} data from a run with
$k_{\rm f}=27$ and compare spectra in the linear and
nonlinear regimes. In the linear regime, all spectra are just shifted
along the ordinate, so the spectra have been compensated by the
factor $M_{\rm ini}\exp(2\lambda t)$, where $\lambda$ is the growth
rate and $M_{\rm ini}$ the initial magnetic energy. In the nonlinear
regime the bump on the right stays at approximately the same wavenumber
(the forcing wavenumber), while the bump on the left propagates gradually
further to the left. As it does so, and since the amplitude of the
secondary peak even increases slightly, the net magnetic helicity
inevitably increases (or rather becomes more negative in the present
case). But because of the asymptotic magnetic helicity conservation,
this can only happen on a slow resistive time scale. This leads to
the appearance of a (resistively) slow saturation phase past the
initial saturation; see \Fig{Fpjbm_decay_nfit_run3passot}.

\subsection{The final saturation value}

In a periodic box with helically driven turbulence, the final
saturation value of the magnetic field is determined
by the ratio of the size of the domain to the scale of the forcing.
This is best seen by considering the magnetic helicity equation
\EQ
{\dd\over\dd t}\bra{\AAA\cdot\BB}=-2\eta\mu_0\bra{\JJ\cdot\BB}.
\label{hel_cons}
\EN
If $\bra{\AAA\cdot\BB}$ were not gauge invariant (for example if there
are open boundaries), \Eq{hel_cons} would be useless. In particular,
$\bra{\AAA\cdot\BB}$ will in general not be constant in the steady state
(see Fig.~2 of Ref.~\cite{BDS02}, for an example).
One therefore has to go to the gauge-invariant relative magnetic
helicity of Berger \& Field \cite{BF84}. This has been done in
Refs~\cite{BD01,BDS02}.
In the present case of periodic boundaries, however, $\bra{\AAA\cdot\BB}$
is automatically gauge invariant and therefore a physically meaningful quantity, so
it must be constant in the steady state. \EEq{hel_cons} says that then
the current helicity, $\bra{\JJ\cdot\BB}$, must vanish. At first
glance this seems to be in conflict with the idea of building up a
helical large scale field. The solution is that there must then be
an equal amount of small scale current helicity so that
\EQ
\bra{\JJ\cdot\BB}=\bra{\meanJJ\cdot\meanBB}+\bra{\jj\cdot\bb}=0
\quad\mbox{(in the steady state)}.
\label{JdotB=0}
\EN
If the field is fully helical, we have
$\mu_0\bra{\meanJJ\cdot\meanBB}=\mp k_1\bra{\meanBB^2}$ and
$\mu_0\bra{\jj\cdot\bb}=\pm k_{\rm f}\bra{\bb^2}$, where the
upper (lower) sign applies to the case where the small scale
helicity at the forcing scale is positive (negative). We then
have from \Eq{JdotB=0}
\EQ
\bra{\meanBB^2}=(k_{\rm f}/k_1)\,\bra{\bb^2}.
\label{meanBB2final}
\EN
To a good approximation, $\bra{\bb^2}^{1/2}$ will be close to the
equipartition field strength, $B_{\rm eq}$, so
\EQ
\bra{\meanBB^2}/B_{\rm eq}^2\approx k_{\rm f}/k_1>1.
\label{BBeq}
\EN
This means that the energy of the large scale field must, in the final
state, be in super-equipartition by a factor approximately equal
to the scale separation. We recall that this applies to the case of
periodic boundaries. For closed (e.g.\ perfectly conducting) boundaries,
$\bra{\meanBB^2}$ can be even larger than $k_{\rm f}/k_1$ times the
equipartition value \cite{BDS02}. This is because the large scale field
is no longer fully helical, while the small scale field still is. In the
presence of {\it open} boundaries, on the other hand, the large scale
field will generally be smaller than suggested by \Eq{BBeq}, unless the
boundaries transmit preferentially small scale fields (\Sec{Sopen}).

\subsection{Sensitivity to using hyperdiffusivity}

The above statements can readily be generalized to the case where
the usual magnetic diffusion operator, $\eta\nabla^2$, is
replaced by hyperdiffusion, $(-1)^{n-1}\eta_n\nabla^{2n}$ with $n=2$.
This implies that the diffusion has now become more strongly
wavenumber dependent; from $\eta k^2$ to $\eta_2 k^4$. If the diffusion
is the same at small scales, then the diffusion at large scales must
be significantly smaller in the hyperdiffusive runs. This leads to a
dramatic {\it increase} of the resistive saturation time. At the same time
the final saturation field strength is no longer given by \Eq{BBeq}.
The rate of magnetic helicity dissipation is now no longer
proportional to $k$, but to $k^3$. Therefore the final saturation
field strength is given by
\EQ
\bra{\meanBB^2}/B_{\rm eq}^2\approx(k_{\rm f}/k_1)^3\gg1.
\label{BBeq_hyp}
\EN
This result was confirmed also numerically \cite{BS02}. Again, this
applies to periodic boundaries. Hyperdiffusion has been used in the
past in connection with open boundaries \cite{BNST95,GR95}, but it is
not clear how serious the possible artifacts from hyperdiffusion would
be in such cases with open boundaries.

\subsection{The magnetic helicity constraint}
\label{Smaghelconst}

The case of periodic boundaries is particularly useful as a benchmark
to all dynamos exhibiting large scale dynamo action due to the helicity
effect. Here we discuss the functional form $\meanBB^2(t)$ for the late
saturation phase.

\EEq{JdotB=0} allows us not only to determine the final saturation
strength, but also the approximate time evolution near saturation.
As before, we make the assumption of fully helical fields. However,
given that prior to saturation
$|\bra{\jj\cdot\bb}|\approx|\bra{\meanJJ\cdot\meanBB}|$, we must
have $|\bra{\aaaa\cdot\bb}|\ll|\bra{\meanAA\cdot\meanBB}|$, so we
can set
\EQ
\bra{\AAA\cdot\BB}\approx\bra{\meanAA\cdot\meanBB}
=\mp k_1^{-1}\bra{\meanBB^2},
\EN
where the upper (lower) sign refers to positive (negative)
small scale (kinetic and magnetic)
helicity. Inserting this into \Eq{hel_cons} we have
\EQ
k_1^{-1}{\dd\over\dd t}\bra{\meanBB^2}=-2\eta k_1\bra{\meanBB^2}
+2\eta k_{\rm f}\bra{\bb^2}.
\label{meanBBevol}
\EN
The small scale field saturates first, so \Eq{meanBBevol} can then
be integrated to get the {\it subsequent} evolution of $\bra{\meanBB^2}$
toward full saturation \cite{B01}
\EQ
\bra{\meanBB^2}
={k_{\rm f}\over k_1}\bra{\bb^2}
\left[1-e^{-2\eta k_1^2(t-t_{\rm sat})}\right]
\quad\mbox{(for $t>t_{\rm sat}$)},
\label{helconstraint}
\EN
where $t_{\rm sat}$ is the time when the small scale field has reached saturation.
\EEq{helconstraint} is what we usually mean by the {\it magnetic helicity
constraint}.

\subsection{Inverse cascade versus $\alpha$-effect}

The process outlined above can be interpreted in two different ways: inverse
cascade of magnetic helicity and/or $\alpha$-effect. The two are similar in that
they tend to produce magnetic energy at scales larger than the energy-carrying
scale of the turbulence. As can be seen from
\Figs{Fpspec_growth_passot}{Fpspec_pm_satkin}, the present simulations support
the notion of {\it nonlocal} inverse transfer \cite{B01}. This is not really an
inverse cascade in the usual sense, because there is no sustained flux of
energy through wavenumber space as in the direct Kolmogorov cascade. Instead,
there is just a bump traveling to smaller $k$ in wavenumber space. In that
respect, the present simulations seem to differ from the Eddy Damped
Quasi-Normal Markovian (EDQNM) closure approximation \cite{PFL76}.

The other interpretation is in terms of the $\alpha$-effect. We recall that
there is a wavenumber $k_{\max}$ where the growth of the large scale field
is largest; see \Eq{kmax}. For reasonable estimates,
$k_{\max}$ coincides with the position of the secondary bump in the
spectrum (Ref.~\cite{B01}, Sect.~3.5). This can be taken as evidence in favor of
the $\alpha$-effect. In the nonlinear regime, the secondary bump travels
to the left in the spectrum (i.e.\ toward smaller $k$). In the
EDQNM picture this has to do with the equilibration of kinetic and
current helicities at progressively smaller wavenumbers, which then
saturates further growth at that wavenumber, but permits further
growth at smaller wavenumbers until equilibration occurs.
Another interpretation is simply that if $\alpha$ is quenched
to a smaller value, $k_{\max}=\alpha/(2\eta_{\rm T})$ peaks at smaller
wavenumbers where the growth has not yet saturated, until equilibration
is attained also at that scale.

\subsection{Implications}

The growth of the large scale magnetic field can be interpreted as being
due to the $\alpha$-effect. Consequently, a slow-down in the
final saturation phase must have to do with a suppression of
$\alpha$. According to closure models \cite{PFL76}, the $\alpha$-effect
is really the residual between two competing effects: a kinematic
helicity effect (which itself decreases somewhat for strong magnetic
fields), and a current helicity effect of opposite sign.
In \Sec{SDynQuen} we shall use this phenomenology in a mean-field
model where an explicitly time-dependent equation for the current
helicity is solved. It turns out that in this model the late
saturation phase is resistively limited--just like in the simulations.

It is somewhat worrisome that in the nonlinear regime the value
of $\alpha$ depends on the microscopic magnetic diffusivity, which is
very small in most astrophysical situations. One might therefore be
concerned that in astrophysical dynamos the saturation would be
unacceptably slow. In order to say more about
$\alpha$ and also the turbulent magnetic diffusivity $\eta_{\rm t}$, we
need to determine how $\alpha$ and $\eta_{\rm t}$ depend on $\BB$. This
can either be done directly \cite{BrSo02},
which is difficult and the results are noisy,
or we can compare with models that incorporate the effects of $\alpha$
and $\eta_{\rm t}$ quenching. Before we do this (\Secs{Squenching}{SDynQuen}),
we first want to assess the effects of boundaries.

\section{Open boundaries: good or bad?}
\label{Sopen}

\begin{figure}[h!]\centering\includegraphics[width=0.95\textwidth]{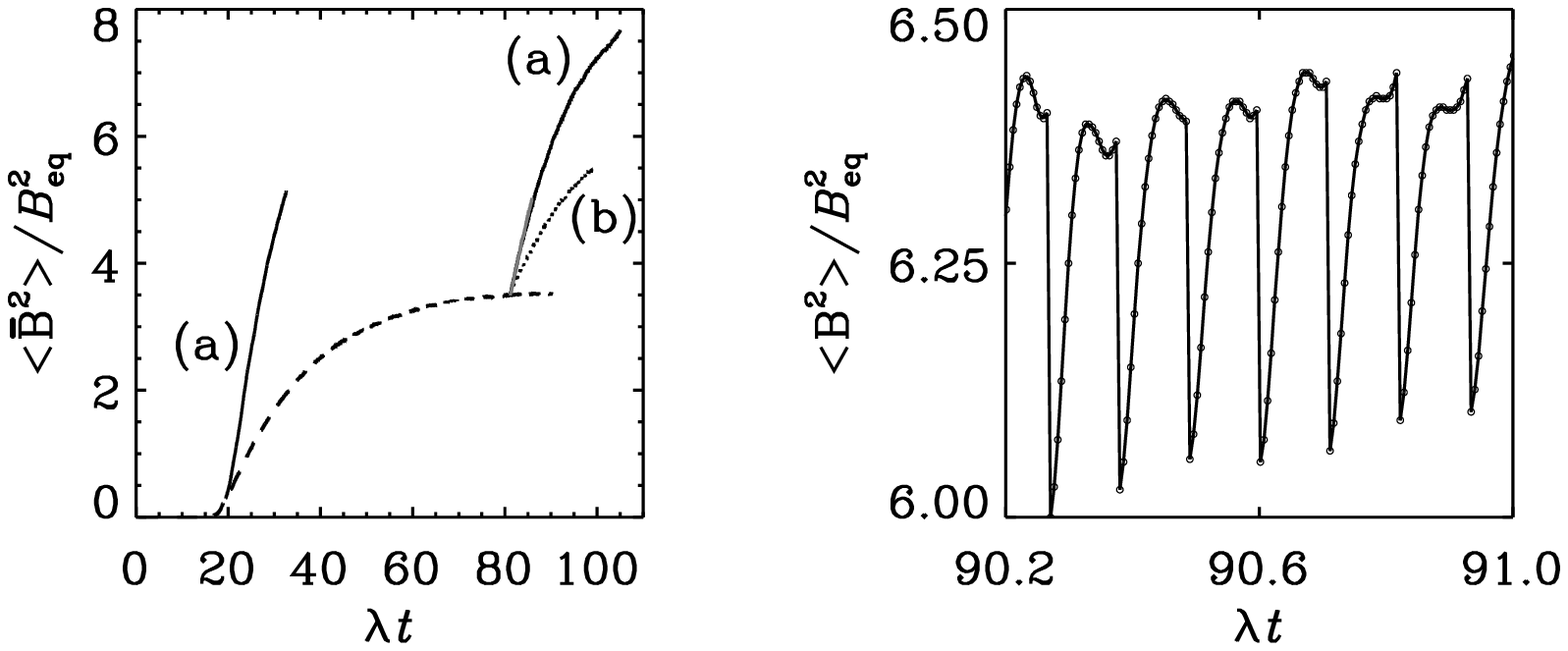}\caption{
The effect of removing small scale magnetic energy in regular time intervals
$\Delta t$ on the evolution of the large scale field (solid lines).
The dashed line gives the evolution of $\bra{\meanBB^2}$ for Run~3 of Ref.~\cite{B01}
(where no such energy removal was included) in units of
$B_{\rm eq}^2=\mu_0\rho_0\bra{\uu^2}$. The two solid lines show the evolution of
$\bra{\meanBB^2}$ after restarting the simulation from Run~3 of Ref.~\cite{B01}
at $\lambda t=20$ and $\lambda t=80$. Time is scaled with the kinematic
growth rate $\lambda$. The curves labeled (a) give the result for
$\Delta t=0.12\lambda^{-1}$ and those labeled (b) for $\Delta t=0.4\lambda^{-1}$.
The second panel shows, for a short time interval, the
sudden drop and subsequent recovery of the total (small and large scale)
magnetic energy in regular time intervals. (Adapted from Ref.~\cite{BDS02}.)
}\label{Fpbmean}\end{figure}

Boundaries generally lead to a loss of magnetic field both on small and
large scales. Losses at large scale tend to lower the saturation field
strength of the mean field. Losses of small scale fields can, at least
in principle, enhance the large scale field \cite{BF00,KMRS00}. This
has been demonstrated in an idealized numerical experiment \cite{BDS02},
where the magnetic field at the forcing wavenumber and beyond had been
removed in regular time intervals; see \Fig{Fpbmean}.
We discuss this now in more detail.

\subsection{Enhancement through losses at small scales}

The enhancement of large scale field by losses at small scales may seem
somewhat mysterious. One way to interpret this result is by saying that
the slow growth occurred because the energy of the small scale magnetic
field has already reached the level of the kinetic energy, and only the
large scale field has not yet saturated. After small scale magnetic
fields have been removed, the field is for a short time interval in
sub-equipartition at small scales and so the overall field (both at small
and large scales) can then grow further during the short time interval
during which the small scale field has not yet fully recovered to the
equipartition value. The effect of a single such
event is small, but many such events can produce a significant effect.
This is exactly what is seen. Another way of interpreting this
result is in terms of mean-field theory where the $\alpha$-effect
is saturated by a cancelation of kinetic and current helicities.
If small scale magnetic fields are removed, the residual $\alpha$-effect
can be larger for some time interval, which then allows the field
to grow further.
In the following we illuminate this result further by considering
a modified magnetic helicity constraint for the case of open boundaries.

\subsection{The modified magnetic helicity constraint}

In \Sec{Smaghelconst} we have discussed an equation for the evolution of
the magnetic energy of the large scale field at late times. Here we have
assumed that there is no loss of magnetic energy and magnetic helicity
through the boundaries. This equation has been generalized to account
for losses of {\it large scale} magnetic helicity \cite{BD01,BDS02}. The
idea is that there will be a magnetic helicity flux that is proportional
to the gradient of the large scale magnetic helicity density (in a fixed
gauge), and hence to the gradient of the magnetic energy density. This
gives rise to an extra diffusion term, and hence to a renormalized,
{\it effective} magnetic diffusivity, $\eta_{\rm eff}^{(1)}$, i.e.\ the
term $2\eta k_1\bra{\meanBB^2}$ has to be replaced by
$2\eta_{\rm eff}^{(1)}k_1\bra{\meanBB^2}$. Therefore, \Eq{meanBBevol}
takes the form \cite{BDS02}
\EQ
k_1^{-1}{\dd\over\dd t}\bra{\meanBB^2}=
-2\eta_{\rm eff}^{(1)}k_1\bra{\meanBB^2}+2\eta k_{\rm f}\bra{\bb^2},
\label{meanBBevol1}
\EN
which has the solution
\EQ
\bra{\meanBB^2}
={\eta k_{\rm f}\over\eta_{\rm eff}^{(1)}k_1}\bra{\bb^2}
\left[1-e^{-2\eta_{\rm eff}^{(1)}k_1^2(t-t_{\rm sat})}\right]
\quad\mbox{(for $t>t_{\rm sat}$)},
\label{helconstraint1}
\EN
Note that the saturation amplitude is decreased by a factor
$\eta/\eta_{\rm eff}^{(1)}$ compared with \Eq{helconstraint}, but
at the same time the $e$-folding time has decreased to
$(2\eta_{\rm eff}^{(1)}k_1^2)^{-1}$.
We return to this behavior in the next subsection.

When the losses through the surface occur preferentially at small scales,
an effective diffusivity would instead affect the small scale helicity
flux. Therefore, \Eq{helconstraint} takes then the form
\EQ
k_1^{-1}{\dd\over\dd t}\bra{\meanBB^2}=
-2\eta k_1\bra{\meanBB^2}
+2\eta_{\rm eff}^{\rm(f)} k_{\rm f}\bra{\bb^2},
\label{meanBBevolf}
\EN
which has the solution
\EQ
\bra{\meanBB^2}
={\eta_{\rm eff}^{\rm(f)} k_{\rm f}\over\eta k_1}\bra{\bb^2}
\left[1-e^{-2\eta k_1^2(t-t_{\rm sat})}\right]
\quad\mbox{(for $t>t_{\rm sat}$)},
\label{helconstraintf}
\EN
Note that the saturation amplitude is now increased by a factor
$\eta_{\rm eff}^{\rm(f)}/\eta$ compared with \Eq{helconstraint},
but the $e$-folding time, $(2\eta k_1^2)^{-1}$,
is still resistively limited. This is
in good with what is seen in the simulations; see \Fig{Fpbmean}.
In reality, there will be both small and large scale losses, so the
large scale magnetic energy is expected to evolve according to
\EQ
\bra{\meanBB^2}={\eta_{\rm eff}^{\rm(f)} k_{\rm f}
\over\eta_{\rm eff}^{(1)} k_1}\bra{\bb^2}
\left[1-e^{-2\eta_{\rm eff}^{(1)}k_1^2(t-t_{\rm sat})}\right]
\quad\mbox{(for $t>t_{\rm sat}$)}.
\label{helconstraintf1}
\EN
This equation allows time scales and saturation amplitudes that
are not resistively limited.

\subsection{Simulations with open boundaries}

\begin{figure}[h!]\centering\includegraphics[width=0.95\textwidth]{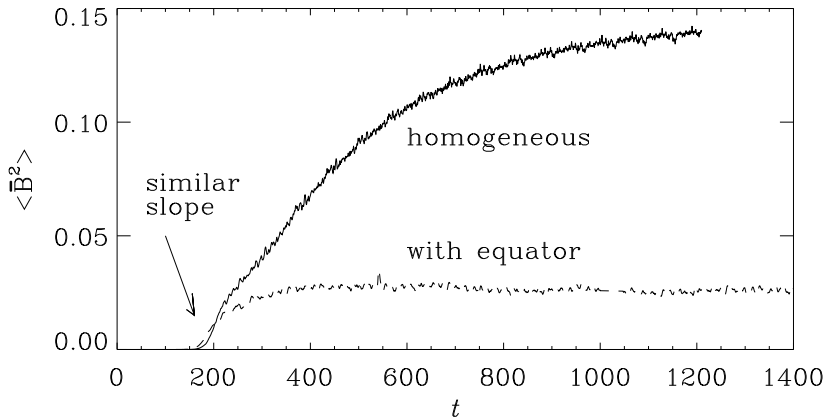}\caption{
Evolution of the magnetic energy for a run with homogeneous forcing
function (solid line) and a forcing function whose helicity varies
sinusoidally throughout the domain (dotted line) simulating the effects
of equators at the two nodes of the sinusoidal helicity profile.
}\label{Fpbmean_Run2}\end{figure}

So far, simulations have not yet shown that the losses of small scale
magnetic fields are actually stronger than those of large scales fields.
Simulations with a vertical field (pseudo-vacuum) boundary condition
have shown that most of the magnetic energy is lost at small scales
\cite{BD01}. The way this affects the slow resistively limited saturation
process discussed earlier is by simply cutting off the saturation process
at an earlier time, without changing the approximately linear slope
past the initial saturation; cf.\ \Eq{helconstraint1}.
In \Fig{Fpbmean_Run2} we demonstrate
a very similar behavior in another system which is actually periodic,
but the helicity of the forcing is modulated in the $z$-direction
such that the sign of the kinetic helicity changes in the middle.
One can therefore view this system as two subsystems with a boundary
in between. This boundary would correspond to the equator in a star
or the midplane in a disc.

\begin{figure}[h!]\centering\includegraphics[width=0.95\textwidth]{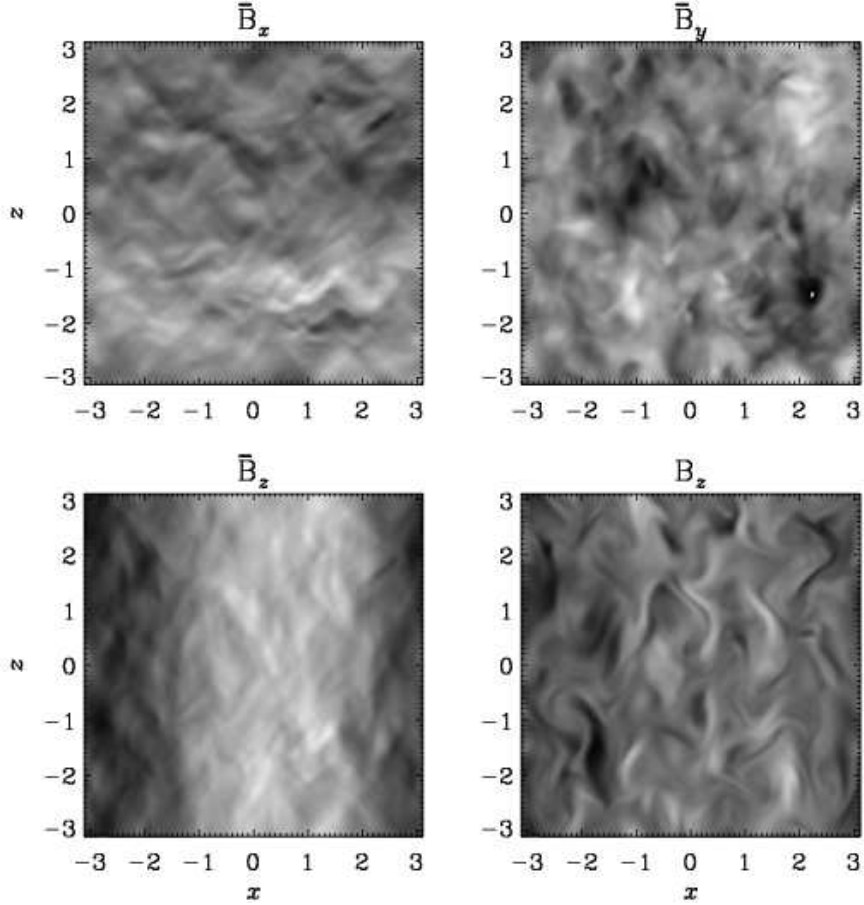}\caption{
Images of the three components of the mean field (averaged over the $y$-direction)
for a run with sinusoidally varying helicity in the $z$-direction.
Note that the most pronounced component of the mean field is
actually $\overline{B}_z(x)$. The large scale field is also
visible in a $y$-slice of $\overline{B}_z$ (last panel).
}\label{Fpbhel}\end{figure}

\begin{figure}[h!]\centering\includegraphics[width=0.95\textwidth]{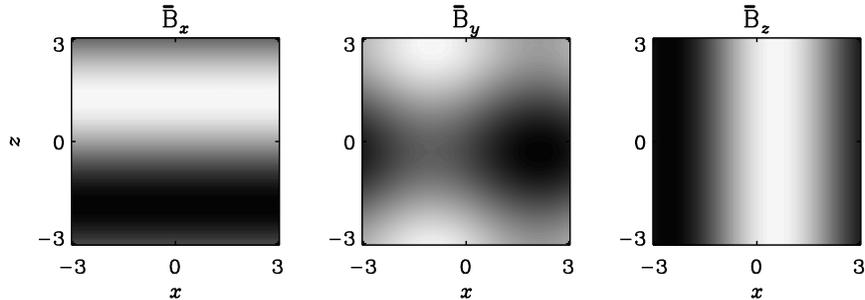}\caption{
Images of the three components of the mean field for an $\alpha^2$ dynamo
with sinusoidally varying $\alpha$-effect in the $z$-direction.
The $\meanB_x$ and $\meanB_z$ components resemble those in
the direct simulation shown in \Fig{Fpbhel}.
}\label{Fpbhel_model}\end{figure}

A somewhat surprising property of the models with variation of helicity
in the $z$-direction is the fact that the mean field varies mostly in
the $x$-direction, i.e.\ it follows the variation of the background model
only weakly; see \Fig{Fpbhel}. Therefore, the mean field must be allowed
to be two-dimensional, i.e.\
\EQ
\meanBB(x,z,t)={\textstyle\int}\BB\dd y\left/{\textstyle\int}\dd y.\right.
\EN
Similar behavior was also found in simulations with boundaries, especially when
the aspect ratio was large \cite{BDS02}. In the present context we were
able to confirm, using a two-dimensional mean field dynamo in periodic
geometry, that for $\alpha\propto\sin k_1z$ the most easily excited mode
varies indeed both in $x$ and $z$; see \Fig{Fpbhel_model}.

In the simulations presented so far, boundaries merely tend to
{\it reduce} the final saturation field strength. Thus, the idea
to enhance the large scale field by small scale losses is not
currently supported by simulations. It is quite possible, however, that this is
simply a consequence of too simple a representation of the physical
boundary. In the sun, coronal mass ejections are quite vigorous
events that are known to shed large amounts of helical magnetic fields
\cite{BR00,DeV00,Chae00,Low01}. This kind of physics is
not at all represented by adopting vacuum
or pseudo-vacuum (vertical field) boundary conditions, as was done
in Ref.~\cite{BDS02}.

We may then conclude that in simulations of large scale dynamos with
relatively simple boundary conditions, open boundaries tend to be more
important for large scale fields than for small scale fields. Although
more realistic boundary conditions still need to be considered, it
is useful to study more carefully whether, on observational grounds,
a resistively limited dynamo can indeed be clearly excluded.

\section{How long is long? -- or what the skin depth has to do with the solar cycle}

In this section we want to estimate the amount of magnetic helicity that is to be
expected for a model of the solar dynamo. We also need to know which
fraction of the magnetic field takes part in the 11-year cycle.
Following an approach similar to that of Berger \cite{Ber84}, we
can bound the rate of change of magnetic helicity in terms of the rate
of Joule dissipation, $Q_{\rm Joule}$, and magnetic energy, $M$.
For an oscillatory dynamo, all three
variables, $H$, $M$, and $Q_{\rm Joule}$ vary in an oscillatory
fashion with a cycle frequency $\omega$ of magnetic energy
(corresponding to 11 years for the sun -- not 22 years), so
we estimate
$|\dd H/\dd t|\la\omega|H|$ and $Q_{\rm Joule}\la\omega M$, which
leads to the inequality \cite{BDS02,BS02}
\EQ
|H|/(2\mu_0 M)\leq\ell_{\rm skin},
\label{H-M-skin}
\EN
where $\ell_{\rm skin} = (2\eta/\omega)^{1/2}$ is the skin depth,
here associated with the oscillation frequency $\omega$.
Thus, the maximum magnetic helicity that
can be generated and dissipated during one cycle is characterized
by the length scale $|H|/(2\mu_0 M)$, which has to be less than
the skin depth $\ell_{\rm skin}$.

For $\eta$ we have to use the Spitzer resistivity which is
proportional to $T^{-3/2}$ ($T$ is temperature), so $\eta$ varies between
$10^4\cm^2/{\rm s}$ at the base of the convection zone to about $10^7\cm^2/\s$
near the surface layers and decreases again in the solar atmosphere.
Using $\omega=2\pi/(11\yr)=2\times10^{-8}\s^{-1}$
for the relevant frequency at which $H$ and $M$ vary we have
$\ell_{\rm skin}\approx10\km$ at the bottom of the convection zone
and $\ell_{\rm skin}\approx300\km$ at the top.

This needs to be compared with the value $|H|/(2\mu_0 M)$ obtained from
dynamo models. Although mean-field theory has been around for several
decades, the helicity aspect has only recently attracted significant
attention. In the proceedings of a meeting devoted specifically to this
topic \cite{BCP99}, magnetic helicity was discussed extensively also in
the context of mean-field theory. However, the precise amount of magnetic
helicity relative to the magnetic energy, and the possibility of helicity
reversals at some length scale were not addressed at the time, although
the the evolution of the current helicity has already been investigated
in the context of a mean field model \cite{DikGil01}.

For a sphere (or a half-sphere) with open boundary conditions and volume $V$
(for example the northern hemisphere), one has to use the gauge-invariant
relative magnetic helicity of Berger \& Field \cite{BF84},
\EQ
H=\int_V(\AAA+\AAA_{\rm P})\cdot(\BB-\BB_{\rm P})\,\dd V,
\EN
where $\BB_{\rm P}=\nab\times\AAA_{\rm P}$ is a potential field used as
reference field that has on the boundaries the same normal component
as $\BB$.  Any additional gradient terms, $\nab\phi$, in $\AAA$ or
$\AAA_{\rm P}$ yield only a surface term,
\EQ
\int_{\partial V}\phi(\BB-\BB_{\rm P})\cdot\dd\SSS,
\EN
which vanishes because $\BB_{\rm P}\cdot\nn=\BB\cdot\nn$.
When applied to an axisymmetric mean field, which can be written as
$\meanBB=b\vec{\hat{\phi}}+\nab\times a\vec{\hat{\phi}}$, it turns
out that the relative magnetic helicity integral is simply \cite{BDS02}
\EQ
H=2\int_V ab\,\dd V.
\label{relHm}
\EN
In order to see how the condition (\ref{H-M-skin}) is met by
mean-field models, we have calculated a typical $\alpha\Omega$
model relevant to the sun and evaluated \Eq{relHm} over the volume
of the northern hemisphere \cite{BDS02}.

We recall that in the Babcock-Leighton approach it is mainly the latitudinal
differential rotation that enters. We also note that, although the latitudinal
migration could be explained by radial differential rotation, meridional
circulation is in principle able to drive meridional migration even when
the sense of radial differential rotation would otherwise be wrong for driving
meridional migration \cite{KRS01,Dur95,CSD95}.
Therefore, we have considered the simple models in spherical geometry.
The results of such a model show that \cite{BDS02}, once $B_{\rm pole}/B_{\rm belt}$
is in the range consistent with observations, $B_{\rm pole}/B_{\rm belt}
=(1...3)\times10^{-4}$, $H_{\rm N}/(2\mu_0M_{\rm N}R)$ is around
$(2-5)\times10^{-4}$ for models with latitudinal shear. (Here, the subscript
`N' refers to the northern hemisphere.) This confirms
the scaling
\EQ
H_{\rm N}/(2\mu_0M_{\rm N}R)={\cal O}(B_{\rm pol}/B_{\rm tor})
\ga B_{\rm pole}/B_{\rm belt}.
\EN
Given that $R=700\Mm$ this means that
$H_{\rm N}/(2\mu_0M_{\rm N})\approx70...200\km$, which would be comparable
with the value of $\ell_{\rm skin}$ near the upper parts of the solar
convection zone, or for models with only latitudinal shear.

The surprising conclusion is that the amount of mean field
helicity that needs to be generated in order to explain the large
scale solar magnetic fields is so small, that it may be plausible
that microscopic magnetic diffusion could still play a role in the
solar dynamo. In other words, open boundary effects may well be important
for understanding the time scale of the dynamo, but the effect does not
need to be extremely strong.

\section{Connection with $\alpha$-quenching}
\label{Squenching}

Nonlinear helical dynamos in a periodic domain are particularly
simple. They provide therefore an ideal benchmark for models of
$\alpha$-quenching. Before applying models to more complicated situations,
they should pass the test of predicting the right behavior in the simple
case of an $\alpha^2$-dynamo.
By ``right behavior'' we mean that the large scale magnetic field
saturates as seen in \Eq{helconstraint}. While this is already
a relatively stringent test that allows us to eliminate some
earlier models (see \Sec{Sotherquench}), we point out that the
complete time evolution (including the early kinematic exponential
growth phase) can only be described correctly by an explicitly
time-dependent evolution equation for $\alpha$ that ensures that
the magnetic helicity equation is obeyed exactly at all times.

\subsection{The lorentzian quenching formula}

We first assume that both $\alpha$-effect and the turbulent magnetic diffusivity,
$\eta_{\rm t}$, are being affected by the magnetic field. In the first
class of models we assume
\EQ
\alpha=\alpha_0\,q(\meanBB),\quad
\eta_{\rm t}=\eta_{\rm t0}\,q(\meanBB),
\EN
i.e.\, we postulate the existence of an algebraic quenching formula
for both $\alpha$ and $\eta_{\rm t}$. The models that work best are
those with a lorentzian quenching formula,
\EQ
q(\meanBB)={1\over1+a\meanBB^2/B_{\rm eq}^2}
\quad\mbox{(lorentzian formula)},
\label{lorentzian}
\EN
where $a$ is a dimensionless coefficient and $B_{\rm eq}$ is the
equipartition field strength with $B_{\rm eq}^2/\mu_0=\bra{\rho\uu^2}$.

For an $\alpha^2$-dynamo in a periodic box, the coefficient $a$ can
be readily determined. This is because here we actually know the final
saturation field strength; see \Eq{BBeq}. In a mean-field model, on the
other hand, where quenching is the only nonlinearity, the saturation
field strength is proportional to $a^{-1/2}$. In order to work out
the coefficients, we consider the $\alpha^2$ mean-field equation for a
Beltrami field with wavenumber $k_1$. The evolution of the mean-squared
field strength is then governed by \cite{B01}
\EQ
\half{\dd\over\dd t}\ln\bra{\meanBB^2}=
{\alpha_0k_1-\eta_{\rm t0}k_1^2\over1+a\bra{\meanBB^2}/B_{\rm eq}^2}
-\eta k_1^2.
\label{ddtlnB2}
\EN
In the steady state, the right hand side of \Eq{ddtlnB2} has to vanish, so
\EQ
{\alpha_0k_1-\eta_{\rm t0}k_1^2\over1+a\bra{\meanBB^2}/B_{\rm eq}^2}
-\eta k_1^2=0,
\EN
or, using \Eq{BBeq},
\EQ
a={\alpha_0-\eta_{\rm t0}k_1\over\eta k_{\rm f}}
\equiv{\lambda\over\eta k_{\rm f}k_1}
\label{a}
\EN
(see Ref.~\cite{B01}). The main point here is that the parameter
$a$ scales with the magnetic Reynolds number, so it is very large under
astrophysical circumstances. Moreover, if the growth rate
$\lambda$ scales with the inverse turnover time on the forcing scale,
then the $k_{\rm f}$ in the denominator cancels and $a$ is proportional
to the magnetic Reynolds number \cite{CV91,VC92,CH96}.
This is also consistent with the analysis of Ref.~\cite{BB02}.

A consequence of the large value of $a$ is a very slow saturation
phase and not necessarily a very low saturation level, as is usually
believed \cite{VC92}. That the saturation
phase is slow can be seen, for example, by considering the magnitude
of the right hand side of \Eq{ddtlnB2} near saturation, so we put
\EQ
\bra{\meanBB^2}/B_{\rm eq}^2=(1-\epsilon)k_{\rm f}/k_1,
\EN
which means that
we are a fraction $\epsilon$ away from full equipartition. We then have
\EQ
\mbox{rhs of \Eq{ddtlnB2}}={\alpha_0k_1-\eta_{\rm t0}k_1^2\over1+
{\alpha_0k_1-\eta_{\rm t0}k_1^2\over\eta k_{\rm f}}(1-\epsilon)
{k_{\rm f}\over k_1}}-\eta k_1^2.
\EN
Since the field is already strong, the `1+' in the denominator may be
neglected, so we obtain
\EQ
\mbox{rhs of \Eq{ddtlnB2}}=\left[{\alpha_0k_1-\eta_{\rm t0}k_1^2\over
(\alpha_0k_1-\eta_{\rm t0}k_1^2)(1-\epsilon)}-1\right]\eta k_1^2
={\epsilon\over1-\epsilon}\eta k_1^2\approx\epsilon\eta k_1^2.
\EN
The rhs of \Eq{ddtlnB2} can be regarded as a local growth rate if
the field was unchanged. The extrapolated saturation time would
therefore be $(\epsilon\eta k_1^2)^{-1}$, which is more than a
resistive time!

\begin{figure}[h!]\centering\includegraphics[width=0.95\textwidth]{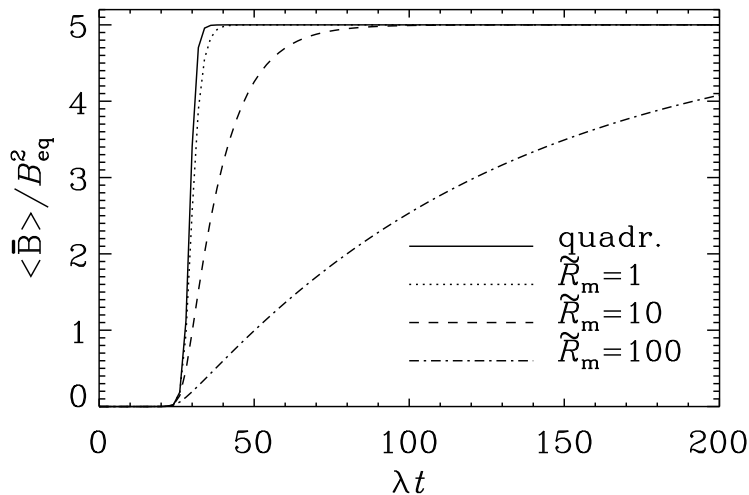}\caption{
Saturation behavior for the quadratic quenching formula (solid line)
compared with different lorentzian quenching formulae with different values
of $\tilde{R}_{\rm m}\equiv\lambda/(\eta k_1^2)$ (broken lines).
Note the slow saturation behavior, consistent with \Eq{helconstraint}.
The results for the quadratic quenching formula are independent of the value of
$\tilde{R}_{\rm m}$, which is therefore inconsistent with \Eq{helconstraint}.
}\label{Fplorqua}\end{figure}

\subsection{Other quenching formulae}
\label{Sotherquench}

By contrast, consider a quenching of the form
\EQ
q(\meanBB)=1-a\meanBB^2/B_{\rm eq}^2
\quad\mbox{(quadratic formula)}.
\label{quadratic}
\EN
To match the right saturation field strength we have to have
\EQ
(\alpha_0k_1-\eta_{\rm t0}k_1^2)(1-ak_{\rm f}/k_1)-\eta k_1^2=0.
\EN
This gives
\EQ
a={\alpha_0k_1-\eta_{\rm T0}k_1^2\over\alpha_0k_1-\eta_{\rm t0}k_1^2}
\,{k_1\over k_{\rm f}}\rightarrow{k_1\over k_{\rm f}}
\quad\mbox{for small $\eta$}.
\EN
Again, we calculate the instantaneous growth rate for a
field that was close to final saturation, i.e.\ we put
$\bra{\meanBB^2}/B_{\rm eq}^2=(1-\epsilon)k_{\rm f}/k_1$.
This gives
\EQ
\mbox{rhs of \Eq{ddtlnB2}}=(\alpha_0k_1-\eta_{\rm t0}k_1^2)
\left[1-{\alpha_0k_1-\eta_{\rm T0}k_1^2\over\alpha_0k_1
-\eta_{\rm t0}k_1^2}(1-\epsilon)\right]-\eta k_1^2
\EN
or, after some simplifications,
\EQ
\mbox{rhs of \Eq{ddtlnB2}}=
\epsilon(\alpha_0k_1-\eta_{\rm T0}k_1^2).
\EN
This means that it will only take a few dynamical time scales
before the extrapolated field will reach final saturation, which
is of course incompatible with the magnetic helicity constraint
\eq{helconstraint}.
\FFig{Fplorqua} shows the great discrepancy between these two
quenching formulae. The quadratic quenching formula gives, for
different values of $\tilde{R}_{\rm m}\equiv\lambda/(\eta k_1^2)$,
always the same saturation behavior, while the lorentzian formula
gives a more prolonged saturation phase as $\tilde{R}_{\rm m}$ is
increased.

\begin{figure}[h!]\centering\includegraphics[width=0.95\textwidth]{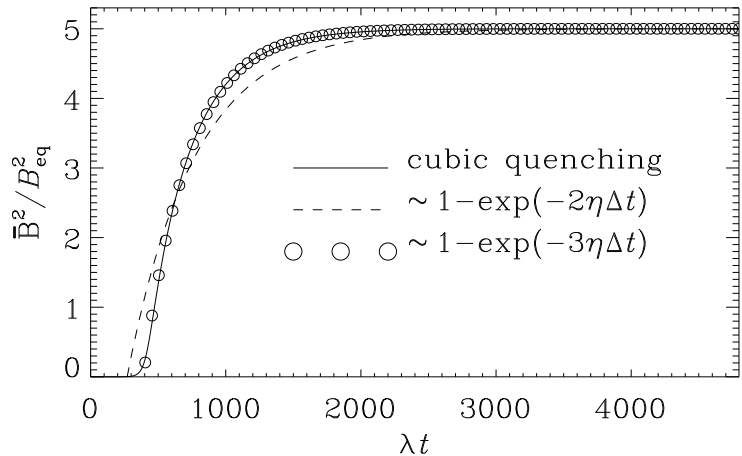}\caption{
Saturation behavior for cubic quenching (solid line) compared with the
magnetic helicity constraint (dashed line). Note also that if we used the
{\it wrong} microscopic diffusivity, $\eta\rightarrow{3\over2}\eta$, the
helicity constraint would actually fit.
}\label{Fpp_cubic}\end{figure}

One may still be tempted to expect that there could be many other
quenching formulae that might work as well. One clear counter example
is shown in \Fig{Fpp_cubic} where we compare the results from a
cubic quenching formula with the magnetic helicity constraint in
\Eq{helconstraint}. For the cubic quenching we just used the very simple
formula $q=1/(1+\beta^3)$, where $\beta^2=a\bra{\meanBB^2}/B_{\rm eq}^2$.
The departure between cubic is not very strong, but clearly noticeable.
We note that if we replaced $\eta\rightarrow{3\over2}\eta$, the
helicity constraint would actually fit, but of course $\eta$ is an input
parameter, so we cannot just adopt a different value in the analysis.
We may therefore conclude that cubic quenching can be ruled out.

\begin{figure}[h!]\centering\includegraphics[width=0.95\textwidth]{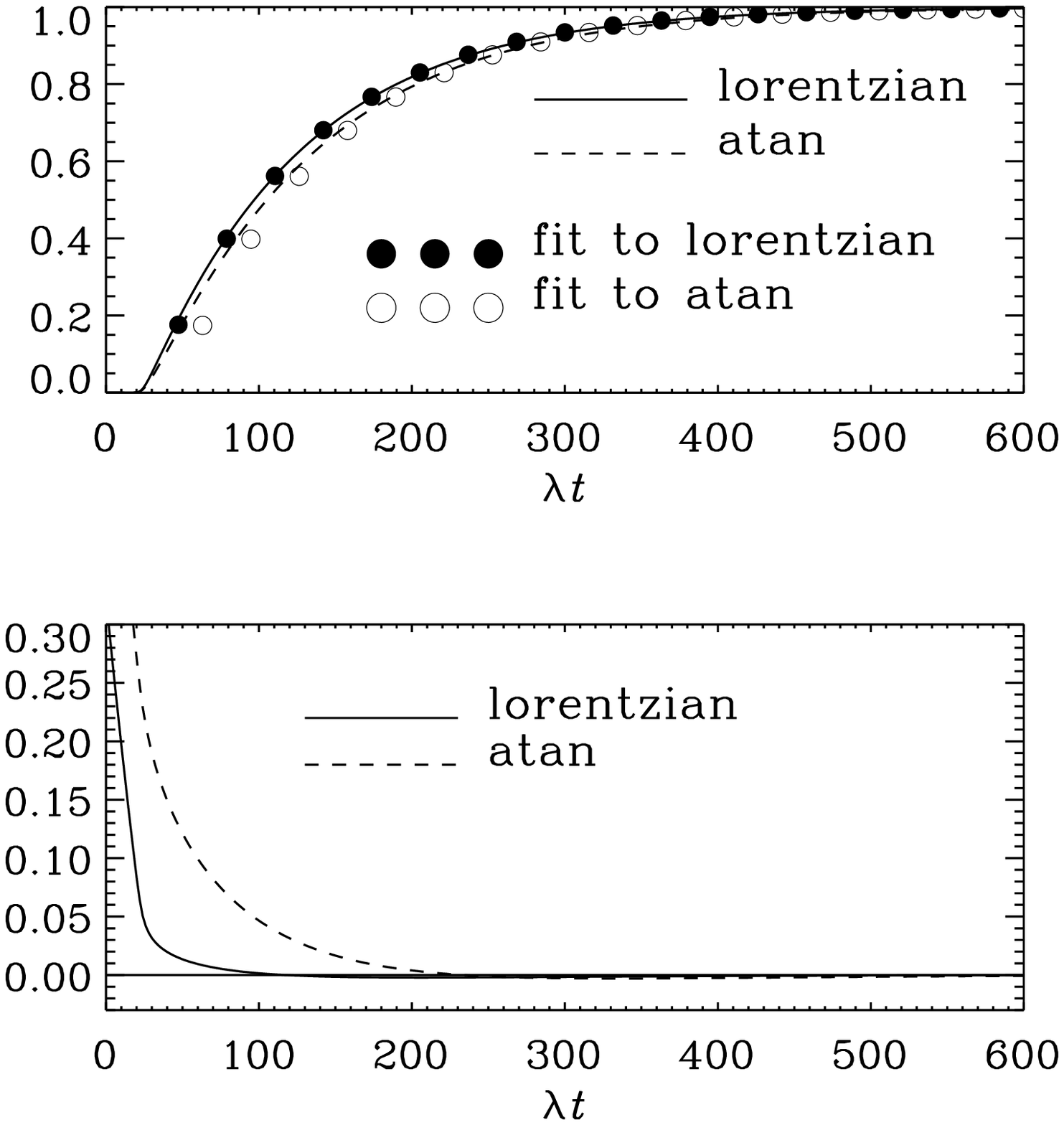}\caption{
Comparison of the saturation behavior for the lorentzian and atan formulae,
together with the corresponding fits obtained from the helicity constraint.
}\label{Fpquench_comp}\end{figure}

Finally, we consider a quenching formula that has the correct strong field
asymptotics, $q\rightarrow\beta^{-2}$, and also the same weak field
Taylor expansion as the lorentzian formula, $q\approx1+\beta^2$,
\EQ
q(\meanBB)={1\over\beta^2}
\left(1-{\tan^{-1}\sqrt{3\beta^2}\over\sqrt{3\beta^2}}\right),
\quad\mbox{where $\;\beta^2=a\bra{\meanBB^2}/B_{\rm eq}^2$}.
\label{atan}
\EN
This expression entered in the quenching formula derived by Field,
Blackman, \& Cho \cite{FBC99}. In \Fig{Fpquench_comp} we compare the saturation
behavior for the atan formula with the lorentzian one. Clearly, the atan
formula is much better than any of the other formulae considered in this
subsection, but the lorentzian formula is still considerably closer to
the magnetic helicity constraint than the atan formula.

\subsection{Non-universality of the lorentzian quenching formula}

We may now be under the impression that the lorentzian formula is probably
the correct quenching expression. While is does indeed provide a good
description of what is going on in the simulations, we note that there
are also a few problems. Firstly, when applied to other models where the
field is in general no longer isotropic and force-free (e.g., if there
is shear or if there are boundaries) the best fit value of $a$ is no longer equal
to the value calculated for the $\alpha^2$-dynamo; see \Eq{a}.

The other problem is rather an uncomfortable prediction from the
quenching model with the lorentzian formula. Since both $\alpha$
and $\eta_{\rm t}$ are quenched by equal amounts, $\eta_{\rm t}$ will
become comparable with $\eta$ near full saturation.
As a consequence, the cycle period of
$\alpha\Omega$-dynamo models becomes comparable to the resistive time
which is rather long. So, mean-field theory of the solar dynamo may face
a very serious problem, unless open boundary effects (\Sec{Sopen}) play
an important role. Simulations perhaps seem to point into this direction
as well: the cycle period found in simulations with shear \cite{BBS01}
was already rather long,
and with a further reduction of the magnetic diffusivity by a factor of
2.5, the cycles disappeared altogether \cite{BDS02}. On the other hand,
the absence of cycles could have been for other reasons, for example due
to too restricted a geometry (with sinusoidal shear flow on a scale only
5 times larger than the energy carrying scale of the turbulence). Thus,
these two simulations are perhaps not yet fully conclusive.

In the following we point out that there is yet another possibility
that is at least equally well in agreement with our $\alpha^2$-dynamo
benchmark result and theoretically more theoretically more appealing
because it satisfies the magnetic helicity equation exactly at all times.

\section{Dynamical quenching}
\label{SDynQuen}

\begin{figure}[t!]\centering\includegraphics[width=0.75\textwidth]{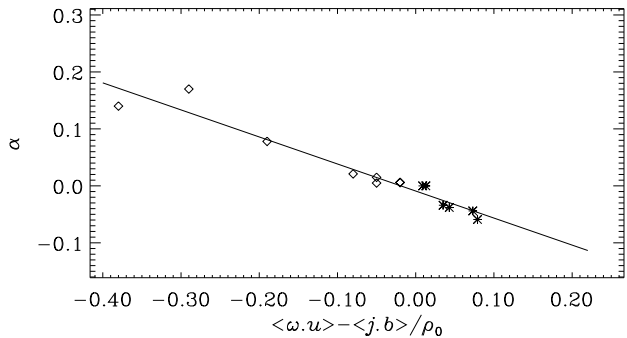}\caption{
Dependence of $\alpha=\overline{\cal E}_y/B_{0y}$ on the residual
helicity, obtained by imposing
a uniform magnetic field $\BB_0=(0,\;B_{0y},\;0)$ and driving the turbulence
either through the momentum equation (diamonds) or by an extra forcing term
in the induction equation (asterisks).
}\label{Fpalphel}\end{figure}

A quenching formula that we have not yet discussed is the formula
for the {\it residual} $\alpha$-effect,
\EQ
\alpha=\alpha_{\rm K}+\alpha_{\rm M}\quad\mbox{with}\quad
\alpha_{\rm K}=-{\tau\over3}\bra{\oo\cdot\uu},\quad
\alpha_{\rm M}=+{\tau\over3\rho_0}\bra{\jj\cdot\bb},
\label{alpM}
\EN
which is due to Pouquet, Frisch, and L\'eorat \cite{PFL76}, and has frequently
been used in connection with $\alpha$-quenching \cite{GD94,GD95,GD96,BY95}.
The result that the $\alpha$-effect is proportional to the {\it residual}
helicity, $\bra{\oo\cdot\uu}-\bra{\jj\cdot\bb}/\rho_0$ has also been
confirmed numerically \cite{Bra99} by imposing a uniform magnetic field
and driving the turbulence either through the momentum equation (as is
done in the rest of the paper) or through a forcing term in the induction
equation; see \Fig{Fpalphel}.

The question now is how to use \Eq{alpM} in a mean field model, which
only knows about the mean field, $\meanBB$. If we were to approximate
$\bra{\jj\cdot\bb}$ in a direct manner by $\meanBB^2$, this would
correspond to the quenching formula \eq{quadratic}, which is clearly
ruled out because it violates the magnetic helicity equation.
On the other hand, we can explicitly make sure that the magnetic
helicity equation \eq{hel_cons} is obeyed. The contribution
$\bra{\meanAA\cdot\meanBB}$ from the large scale fields to the magnetic
helicity equation is automatically taken into account by the mean-field
equation, so we only need to solve for the missing contribution
from small scales, $\bra{\aaaa\cdot\bb}$, and the two equations for
$\bra{\meanAA\cdot\meanBB}$ and $\bra{\aaaa\cdot\bb}$ must be fully
coupled. 

Another way of seeing this \cite{FB02} is that, while $\bra{\AAA\cdot\BB}$
stays close to zero on short enough time scales, any increase of
the magnetic field by the $\alpha$-effect leads to an increase of
the large scale magnetic helicity, $\bra{\meanAA\cdot\meanBB}$. This
can only be consistent with an almost unchanged $\bra{\AAA\cdot\BB}$ if
there is a simultaneous generation of small scale magnetic helicity,
$\bra{\aaaa\cdot\bb}$, of opposite sign, so that
\EQ
\bra{\AAA\cdot\BB}=\bra{\meanAA\cdot\meanBB}+\bra{\aaaa\cdot\bb}
\EN
stays close to zero. The price to pay for this is that the small
scale magnetic helicity can also produce an $\alpha$-effect,
$\alpha_{\rm M}=\onethird\tau\bra{\jj\cdot\bb}/\rho_0$, but it
has the opposite sign than $\alpha_{\rm K}$, so the residual
$\alpha$-effect becomes quenched. Mathematically, this quenching
of $\alpha$ can be described by the magnetic helicity equation.
The contribution of the large scale field to the magnetic
helicity equation follows from the mean-field equation. The small
scale contribution is exactly such that the sum of these two
equations gives \Eq{hel_cons}. Thus, we have a pair of two
equations \cite{FB02}
 \EQ
{\dd\over\dd t}\bra{\meanAA\cdot\meanBB}=
2\bra{\meanemf\cdot\meanBB}-2\eta\mu_0\bra{\meanJJ\cdot\meanBB},
\label{dABdt}
\EN
\EQ
{\dd\over\dd t}\bra{\aaaa\cdot\bb}=
-2\bra{\meanemf\cdot\meanBB}-2\eta\mu_0\bra{\jj\cdot\bb}.
\label{dabdt}
\EN
where $\meanemf=\overline{\uu\times\bb}$ is the mean turbulent
electromotive force, for which we adopt the usual mean-field
closure in terms of $\alpha$-effect and turbulent magnetic
diffusivity, i.e.\
\EQ
\meanemf=\alpha\meanBB-\eta_{\rm t}\mu_0\meanJJ.
\EN
Note that \Eq{dABdt} follows directly from the usual mean-field
dynamo equation \eq{simplest_dyn_eqn}. Making use of the relation
$\mu_0\bra{\jj\cdot\bb}=k_{\rm f}^2\bra{\aaaa\cdot\bb}$ in \Eq{alpM},
\Eq{dabdt} becomes \cite{BB02}
\EQ
{\dd\alpha_{\rm M}\over\dd t}=-2\eta k_{\rm f}^2
\left(R_{\rm m}{\bra{\meanemf\cdot\meanBB}\over B_{\rm eq}^2}
+\alpha_{\rm M}\right),
\label{dynquench}
\EN
where $R_{\rm m}=\eta_{\rm t}/\eta$ is the appropriate definition
in the present context.

An evolution equation for $\alpha$ was already proposed twenty
years ago \cite{KR82}; see also Refs~\cite{ZRS83,KRR95}.
Nevertheless, dynamical quenching
was usually ignored, although it has sometimes been used in mean-field
models with the main motivation to promote and study chaotic behavior in
stellar dynamos \cite{Ruz81,SS91,FJK93,Cea97}. Kleeorin \ea \cite{KMRS00} were the first
to point out that the catastrophic quenching of Vainshtein \& Cattaneo
\cite{VC92} is just a special case of dynamical quenching.

\subsection{Adiabatic approximation and force-free degeneracy}

Near the saturated state the explicit time derivative in \Eq{dynquench},
$\dd\alpha_{\rm M}/\dd t$, can be neglected and the value of $\alpha_{\rm M}$
adjusts `adiabatically' as the field saturates. Thus, we have \cite{BB02}
\EQ
0=R_{\rm m}{\bra{\meanemf\cdot\meanBB}\over B_{\rm eq}^2}+\alpha_{\rm M},
\EN
or, after substituting $\alpha_{\rm M}=\alpha-\alpha_{\rm K}$,
\EQ
R_{\rm m}\left(\alpha\bra{\meanBB^2}-\eta_{\rm t}\bra{\meanJJ\cdot\meanBB}\right)
+\left(\alpha-\alpha_{\rm K}\right)=0,
\EN
which yields
\EQ
\alpha={\alpha_{\rm K}+R_{\rm m}\eta_{\rm t}\bra{\meanJJ\cdot\meanBB}\over
1+R_{\rm m}\bra{\meanBB^2}}.
\label{adiabat_approx}
\EN
This equation was already obtained by Gruzinov \& Diamond \cite{GD94,KRR95}.
The late saturation phase of $\alpha^2$-dynamos is
well described by \Eq{adiabat_approx}. This is because near saturation
time dependence is governed by the slow resistive adjustment
in the mean field equation for the large scale field, whilst the
$\alpha$ equation is quickly adjusting to whatever the large scale
field is at any time.

The reason why the lorentzian quenching formula describes the
resistively limited quenching behavior so well is because in the
case of a nearly force free large scale magnetic field the
lorentzian quenching formula and the adiabatic approximation
become identical.
Indeed, if the large scale magnetic field is force-free, we have
$\bra{\meanJJ\cdot\meanBB}\meanBB=\bra{\meanBB^2}\meanJJ$,
which allows us to write the full electromotive force,
$\meanemf=\alpha\meanBB-\eta_{\rm t}\mu_0\meanJJ$, in the form
\EQA
\meanemf={\alpha_{\rm K}
+R_{\rm m}\eta_{\rm t}\mu_0\bra{\meanJJ\cdot\meanBB}/B_{\rm eq}^2
\over1+R_{\rm m}\bra{\meanBB^2}/B_{\rm eq}^2}\,\meanBB
-\eta_{\rm t0}\mu_0\meanJJ
\nonumber\\
={\alpha_{\rm K}\meanBB
\over 1+R_{\rm m}\bra{\meanBB^2}/B_{\rm eq}^2}
-{\eta_{\rm t0}\mu_0\meanJJ
\over 1+R_{\rm m}\bra{\meanBB^2}/B_{\rm eq}^2},
\label{bothquenched12}
\ENA
which shows that in the force free case the adiabatic approximation
together with constant (unquenched) turbulent magnetic diffusivity becomes
equal to the pair of expressions where both $\alpha$ and $\eta_{\rm t}$
are catastrophically quenched. This is called the force-free degeneracy \cite{BB02}.
This degeneracy is lifted in cases with shear or when the turbulence is
no longer fully helical.

\begin{figure}[t!]\centering\includegraphics[width=0.75\textwidth]{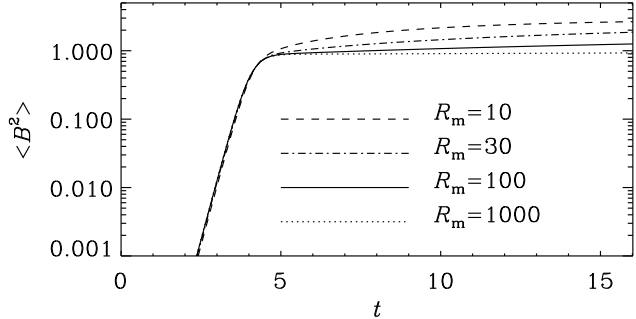}\caption{
The early saturation phase of the energy of the mean field in the
dynamical quenching model for four different values of the magnetic
Reynolds number.
}\label{Fpcomp1}\end{figure}

\begin{figure}[t!]\centering\includegraphics[width=0.75\textwidth]{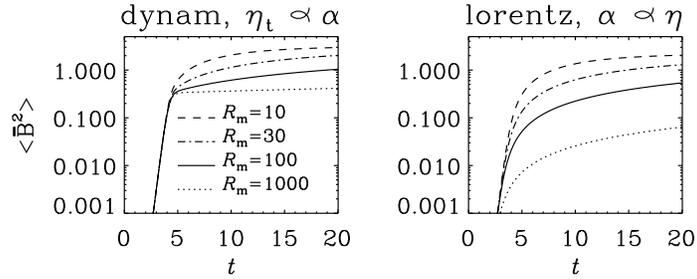}\caption{
The effects of assuming $\eta_{\rm t}\propto\alpha$ in the
dynamical and the lorentzian quenching models for four
different values of the magnetic Reynolds number.
}\label{Fpcomp2}\end{figure}

\begin{figure}[t!]\centering\includegraphics[width=0.75\textwidth]{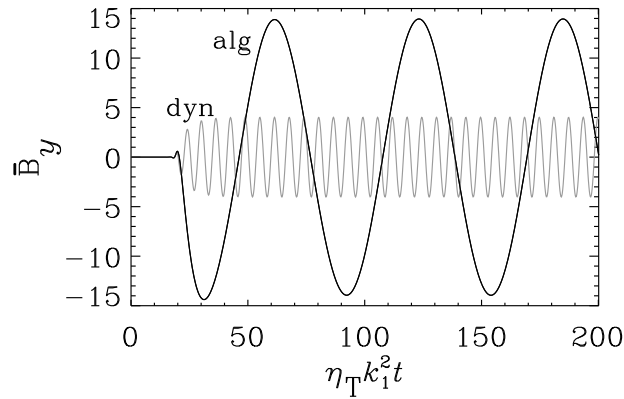}\caption{
Comparison of $\alpha\Omega$-dynamo models with algebraic (solid line)
and dynamical quenching (grey line).
$\Omega'/(\eta_{\rm T}k_1^2)=100$,
$\alpha_{\rm K}/(\eta_{\rm T}k_1)=0.1$,
$R_{\rm m}\equiv\eta_{\rm t}/\eta=10$.
}\label{Fpdynalgom}\end{figure}

\begin{figure}[t!]\centering\includegraphics[width=0.75\textwidth]{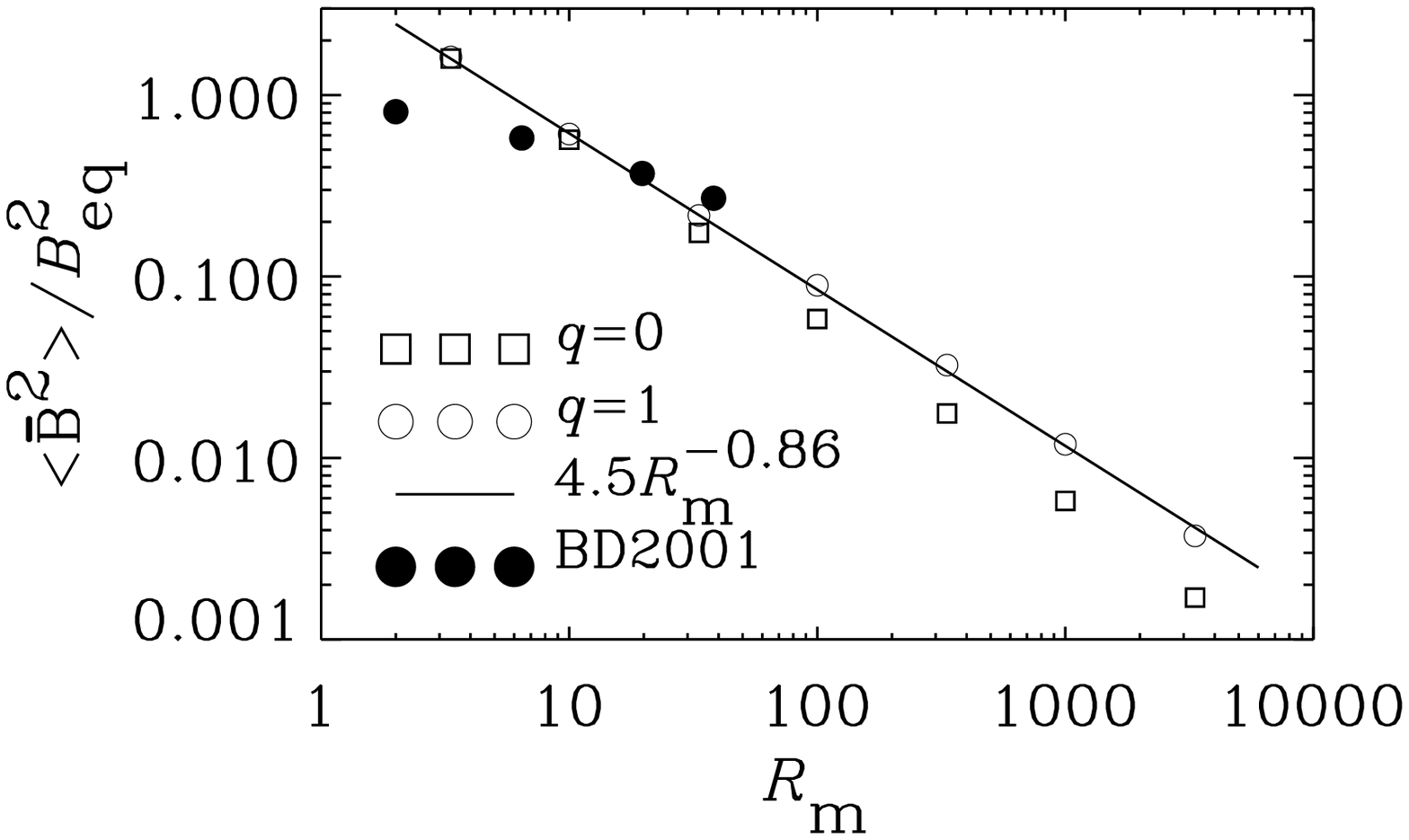}\caption{
Saturation energy versus $R_{\rm m}$ for $\alpha^2$-dynamos with
boundaries.  Models with ($q=1$) and without ($q=0$) loss term are
indicated by open circles and open squares, respectively. The line gives
the fit $\bra{\meanBB^2}\sim R_{\rm m}^{-0.86}$. Simulations of Ref.~\cite{BD01}
are shown as full circles.
}\label{Fpres2}\end{figure}

\subsection{$\alpha^2$-dynamos}

When applied to an $\alpha^2$-dynamo model with dynamical quenching, the
helicity constraint is well satisfied \cite{BB02,FB02} and the
difference between the solutions with dynamical and algebraic quenching
turns out to be small if $R_{\rm m}$ is less than about 1000.
The difference in the evolution of magnetic energy with dynamical
quenching and with algebraic (or lorentzian) quenching,
is only a few percent if the magnetic Reynolds number is small.
The difference increases as the magnetic Reynolds number increases; see
\Fig{Fpcomp2}, where we plot the evolution of $\bra{\meanBB^2}$ around
the time when the kinematic exponential growth turns into the
resistively limited saturation phase which was already described in
\Sec{Smaghelconst}. Conclusive agreement with simulations is at this
point not possible, mostly because the magnetic Reynolds
numbers are not large enough.

\subsection{$\alpha\Omega$-dynamos}

When shear is included, toroidal field can be regenerated solely by the
shear term. This is where dynamical and algebraic quenching lead
to very different behaviors. With algebraic quenching, the reduction of
$\eta_{\rm t}$ leads to resistively long cycle periods in the nonlinear
regime. With dynamical quenching, $\eta_{\rm t}$ is constant and the
cycle frequency remains of order unity; see \Fig{Fpdynalgom}. Here,
$R_{\rm m}$ is only 10, but for larger values the
final field amplitude and the cycle period are considerably enhanced.

\subsection{Open boundaries}

Finally we consider a model with vertical field boundary conditions
($\overline{B}_x=\overline{B}_y=0$
on $z=\pm\pi$) and solve the $\alpha^2$-dynamo equation. The resulting mean-squared
field strength is plotted in \Fig{Fpres2} versus $R_{\rm m}$. We find that
$\bra{\meanBB^2}\sim R_{\rm m}^{-1}$, which is consistent with
\Eq{helconstraint1}, but steeper than what was obtained
in the simulations \cite{BD01}. Kleeorin et al.\ \cite{KMRS00,KMRS02} pointed out that
in \Eq{dynquench} there should be an additional loss term on the right hand side.
By making this loss term suitably $R_{\rm m}$-dependent, one could in principle
make the $R_{\rm m}$-dependence of $\bra{\meanBB^2}$ less steep, but some
reduction is already obtained by allowing for a diffusion-like loss term of the
form $q\eta_{\rm T}\nabla^2\alpha_{\rm M}$ on the right hand side;
see the open circles in \Fig{Fpres2}. The parameter $q$ is used to regulate
the efficiency of this loss. The simulations of Ref.~\cite{BD01}
are shown as full circles.

\begin{figure}[t!]\centering\includegraphics[width=0.95\textwidth]{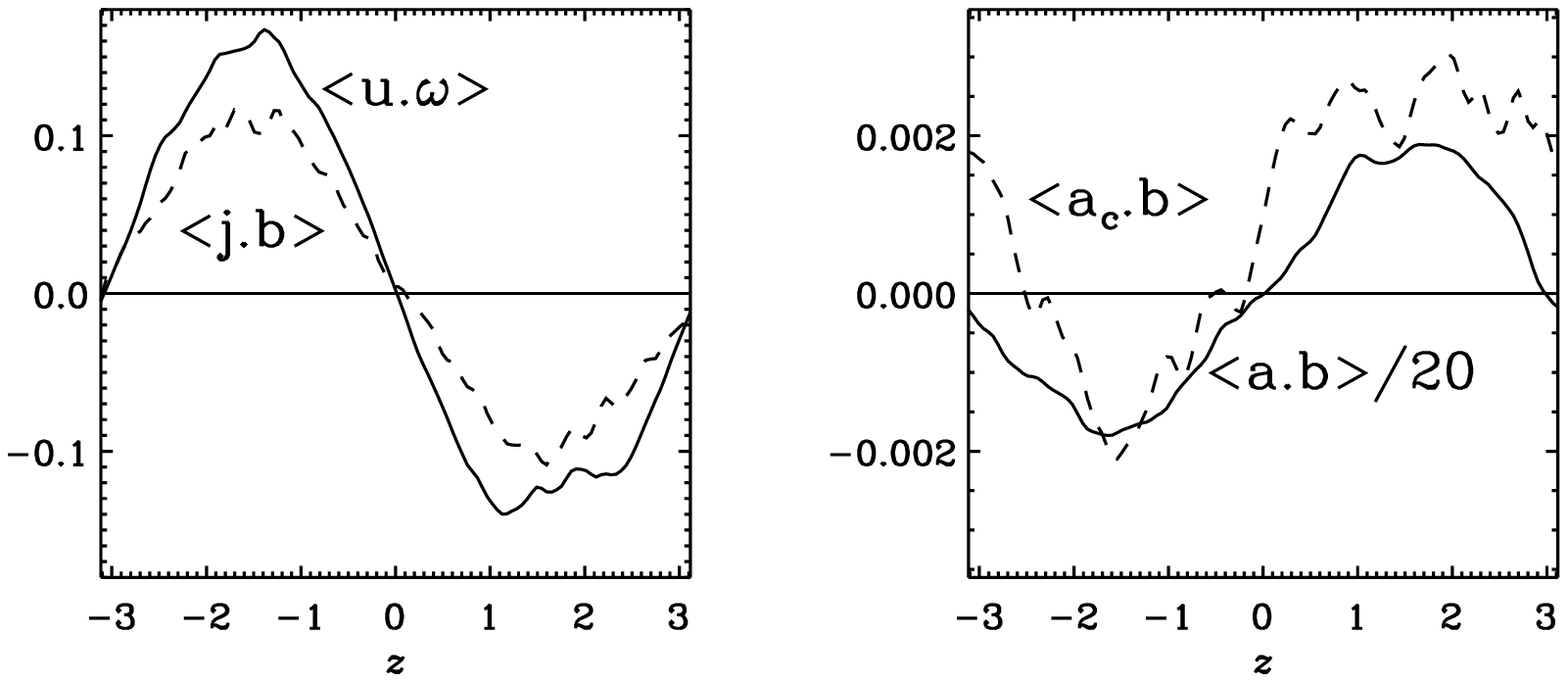}\caption{
Comparison of the horizontally averaged kinetic and current helicity densities,
$\bra{\oo\cdot\uu}$ and $\bra{\jj\cdot\bb}$, respectively (left hand panel),
and the horizontally averaged magnetic helicity densities in the $\phi=0$ gauge 
$\bra{\aaaa\cdot\bb}$ and the Coulomb gauge $\bra{\aaaa_{\rm c}\cdot\bb}$.
}\label{Fphel}\end{figure}

\subsection{Generalization to nonuniform $\alpha$}

In all astrophysical bodies there are opposite signs of kinetic helicity,
$\bra{\oo\cdot\uu}$, in the northern and southern hemispheres. In the
approach of Kleeorin and collaborators \cite{KR82,ZRS83,KRR95} the
dynamical $\alpha$-quenching framework was always used as a theory for
nonuniform $\alpha$, which does not seem to be well justified. Most
importantly, the connection between $\bra{\jj\cdot\bb}$ (in the
expression for $\alpha_{\rm M}$) and $\bra{\aaaa\cdot\bb}$ is no longer
straightforward when the angular brackets denote ensemble averages
(which are not really of practical interest) or averages over one or
two periodic coordinate directions. Indeed, $\bra{\aaaa\cdot\bb}$
is no longer gauge-invariant, so one has to fix the gauge. The
Coulomb gauge is the most common one, but one should realize that
going to another gauge can make a major difference. In \Fig{Fphel}
we show the magnetic helicity in the $\phi=0$ and the Coulomb gauges,
$\bra{\aaaa\cdot\bb}$ and $\bra{\aaaa_{\rm c}\cdot\bb}$, respectively (see
\Sec{Shelical}). Note that the Coulomb gauged magnetic helicity is about
twenty times smaller than that in the $\phi=0$ gauge. More importantly,
the magnetic helicity is {\it not} a positive multiple of the current
helicity, $\bra{\jj\cdot\bb}$. (We recall that for homogeneous turbulence,
$\bra{\jj\cdot\bb}=k_{\rm f}^2\bra{\aaaa\cdot\bb}$.)  Another problem is
that when solving numerically the evolution equation for a space-dependent
$\alpha$-effect one needs for reasons of numerical stability a diffusion
term \cite{KMRS02,Cea98}. However, it is now clear that a loss or exchange
of small scale helicity leads to an enhancement of the large scale field
\cite{BB02}, but from simulations we know that the presence of an equator
rather lowers the energy of the mean field. Perhaps this could be fixed
by adopting an additional loss term in the mean-field equation for the
large scale field, but this procedure would be completely ad hoc. One
may hope however that some kind of a generalization of the dynamical
$\alpha$-quenching is at least in principle possible.

\section{Conclusions}

In this review we have outlined some of the main results of isotropic
MHD simulations in the presence of helicity. We have focussed on
the connection with the $\alpha$-effect in mean-field dynamo theory. We should
emphasize that in the case where the magnetic energy density is uniform
in space, the agreement between simulations
and theory is now well established. In all other cases, things are immediately
more complicated. Moreover, dynamical quenching
cannot readily be generalized to the case where $\alpha_{\rm M}$ varies in space. In
that case the equation for the magnetic helicity density would not
be gauge-invariant. Another problem arises when $\alpha_{\rm K}$ varies in space
and if it changes sign across the equator, for example. These are very
important aspects requiring clarification. It is quite possible
that significant improvement in the theory will soon be possible. Without a corresponding
generalization of dynamic $\alpha$-quenching, if it is ever possible, it would
be difficult to use dynamo theory  in astrophysically interesting circumstances.


%

\end{document}